\documentclass[11pt,a4paper]{article}
\usepackage{jheppub}
\usepackage[english]{babel}
\usepackage{amssymb,amsmath}
\usepackage{mathrsfs}
\usepackage{hyperref}
\usepackage{color}
\usepackage{graphicx}
\usepackage{mathtools}
\usepackage{subfigure}
\usepackage{slashed}
\usepackage{float}
\usepackage{braket}

\DeclarePairedDelimiter\abs{\lvert}{\rvert}%
\newcommand{\zir}{{z_{\rm IR}}}

\title{Holographic Energy Correlators for Soft Walls}

\author[a]{Csaba Cs\'aki,}
\author[a]{Steven Ferrante,}
\author[b]{and Ameen Ismail}
 
\affiliation[a]{Laboratory for Elementary Particle Physics, Cornell University, Ithaca, NY 14853, USA}
\affiliation[b]{Enrico Fermi Institute, University of Chicago, Chicago, IL 60637, USA}

\emailAdd{csaki@cornell.edu}
\emailAdd{sef87@cornell.edu}
\emailAdd{ameenismail@uchicago.edu}

\abstract{ 
We calculate energy correlators in a general holographic model of confinement, involving an asymptotically anti-de Sitter (AdS) warped extra dimension. Building on a recent computation in a minimal hard-wall model of confinement, we show that the shockwave method for efficiently computing energy correlators in AdS generalizes to an arbitrary warped geometry. This is possible because exact, linear shockwave solutions to the 5D field equations exist in any warped background. We apply our formalism to compute the two-point energy correlator for two simple models of confinement with interesting infrared spectra --- one with a gapped continuum spectrum and one with linear Regge trajectories. The results differ from the simple hard-wall model and from each other, demonstrating that the details of the confining dynamics affect the shape of the energy correlator observables.
}

\begin{document}

\maketitle	
\flushbottom

\section{Introduction}

An important outstanding problem in particle physics is to understand the dynamics of confinement in quantum chromodynamics (QCD). This paper concerns the investigation of energy correlator observables as they relate to confinement, which has been the focus of some exciting recent progress~\cite{Hofman_2008,Komiske:2022enw,csaki2024holographicenergycorrelatorsconfining,CMS:2024mlf,CMS-PAS-SMP-22-015,alicecollaboration2024exposingpartonhadrontransitionjets}.

Around the scale $\Lambda_{\rm QCD} \sim 200$~MeV, QCD becomes strongly coupled and quarks are confined into hadrons. The standard tools of perturbation theory cannot grasp the essence of the confinement transition due to its inherently strongly-coupled nature. One can study a strongly-coupled theory on the lattice, of course. Yet, to acquire a deeper theoretical understanding, one can study models that bear some qualitative resemblance to QCD and wherein aspects of confinement are actually calculable. For example, one approach is to study supersymmetric versions of QCD~\cite{Murayama:2021xfj,Csaki:2021jax,Csaki:2023yas,Csaki:2024lvk}. Another idea is to leverage the anti-de Sitter / conformal field theory (AdS/CFT) correspondence~\cite{Maldacena:1997re,Witten:1998qj,Gubser:1998bc} and study 5D models which are holographically dual to 4D theories that resemble QCD in some way (i.e. the AdS/QCD program)~\cite{Sakai:2004cn,deTeramond:2005su,Erlich:2005qh,DaRold:2005mxj,Sakai:2005yt,Karch:2006pv,Csaki:2006ji,Csaki:2008dt,Erlich:2009me}.

As stated above, energy correlators have recently shed new light on confinement. An $n$-point energy correlator is defined as a correlation function of $n$ energy flow operators, $\langle \mathcal{E}(\vec{n}_1) \cdots \mathcal{E}(\vec{n}_n) \rangle$.
Energy flow operators $\mathcal{E}(\vec{n})$ have a simple interpretation in a collider experiment as the energy deposited in a idealized calorimeter placed very far from the interaction point, in the direction specified by $\vec{n}$. This begets an intuitive understanding of an $n$-point energy correlator as a correlation function of the energy deposited in $n$ calorimeters placed at various points on the celestial sphere.

Energy correlators were first studied over four decades ago~\cite{Basham:1977iq,Basham:1978bw,Basham:1978zq,Basham:1979gh}. They have attracted renewed interest in the past 16 years since the seminal work of Hofman and Maldacena studying them in CFTs~\cite{Hofman_2008,Hofman:2009ug}. Hofman and Maldacena also showed that energy correlators may be described using a holographic interpretation. In particular they considered a strongly-coupled 4D CFT which admits a dual description in terms of a weakly-coupled gravitational theory on 5D AdS, described by the metric
\begin{equation}
    ds^2 = \frac{1}{z^2} \left( dx^\mu dx_\mu - dz^2 \right) .
\end{equation}
They showed that energy correlators can be computed  by studying so-called ``shockwave'' deformations of the metric. These are exact solutions to the Einstein field equations which correspond to the insertion of energy flow operators in the dual theory. Since they are exact solutions, performing the computation in this way avoids the need for a laborious perturbative expansion in terms of Witten diagrams.

Since then, the exploration of energy correlators, as well as more general light-ray operators, has spurred developments in our understanding of quantum field theory, e.g.~\cite{Zhiboedov:2013opa,Belitsky:2013xxa,Belitsky:2013bja,Bousso:2015wca,Faulkner:2016mzt,Bousso:2016vlt,Hartman:2016lgu,Casini:2017roe,Balakrishnan:2017bjg,Cordova:2017zej,Cordova:2017dhq,Leichenauer:2018obf,Kravchuk:2018htv,Cordova:2018ygx,Ceyhan:2018zfg,Manenti:2019kbl,Balakrishnan:2019gxl,Belin:2020lsr,Korchemsky:2021htm,Hartman:2023qdn,Hartman:2024xkw}. In tandem with these more formal advances, energy correlators have seen myriad applications in contemporary collider physics~\cite{Dixon:2019uzg,Chen:2019bpb,Chen:2020vvp,Chen:2020adz,Chen:2021gdk,Komiske:2022enw,Holguin:2022epo,Chen:2022jhb,Chen:2022swd,Lee:2022ige,CMS:2024mlf,CMS-PAS-SMP-22-015,alicecollaboration2024exposingpartonhadrontransitionjets,Bossi:2024qho,Lee:2024jnt}. A few years ago~\cite{Komiske:2022enw} showed that the two-point energy correlator provides a beautiful illustration of confinement in QCD, exhibiting a clear transition between two scaling regimes, one characteristic of free hadrons and the other characteristic of asymptotically free quarks and gluons. These scalings can be predicted from first principles~\cite{Lee:2022ige}, but the computation of the transition between the two regimes is infeasible, since it is associated with QCD becoming strongly coupled. See also~\cite{Lee:2024esz,Chen:2024nyc,Liu:2024lxy,Chen:2024nfl} for recent work on hadronization effects in energy correlators.

In light of the vast AdS/QCD literature, a natural question to ask is whether the Hofman--Maldacena computation of energy correlators can be generalized to a holographic model of confinement. Recently~\cite{csaki2024holographicenergycorrelatorsconfining} answered this question in the affirmative, performing the first holographic computation of a two-point energy correlator in a minimal ``hard-wall'' model of confinement. This minimal confinement model is obtained by cutting off AdS with an infrared (IR) brane; the dual interpretation is a 4D CFT which confines at a scale dictated by the brane location. There are interesting complications which arise in going from a gapless to a gapped theory, which were confronted in~\cite{csaki2024holographicenergycorrelatorsconfining}.

The results of~\cite{csaki2024holographicenergycorrelatorsconfining} revealed a beautiful transition from a nearly constant two-point correlator at small angles, which is expected in a strongly-coupled CFT, to an exponentially decaying behavior at larger angles, which is suggestive of confinement. Despite this success, there remain open questions concerning the interplay of holography, energy correlators, and confinement. It is not known whether the calculation performed in~\cite{csaki2024holographicenergycorrelatorsconfining} can be generalized to more sophisticated holographic models of confinement, along the lines of the AdS/QCD program, where the bulk metric is not AdS. Additionally, some details of the transition observed in~\cite{csaki2024holographicenergycorrelatorsconfining} seem qualitatively different from confinement in QCD --- in particular, the exponential decay of the correlator and the manifestation of confinement only in the back-to-back limit. In QCD, confinement is most prominent in the collinear limit, where the EEC approaches zero like an integer power law characteristic of classical, noninteracting hadrons~\cite{CMS-PAS-SMP-22-015}. One would like to know whether these are universal features of holographic confinement or idiosyncrasies of the hard-wall model.

To address these questions the next logical step is clearly to generalize the energy correlator calculation beyond the minimal hard-wall model. That is the goal of this work\footnote{Early versions of some of our results were obtained in~\cite{ameenthesis}.}. Specifically we focus on 5D models with a metric
\begin{equation}
    ds^2 = e^{-2A(z)} \left( dx^\mu dx_\mu - dz^2 \right) ,
\end{equation}
described by a ``warp factor'' function $A(z)$. The behavior of the warp factor in the deep IR, at large $z$, parametrizes the dynamics of confinement in the dual 4D theory. These sort of ``soft-wall'' models of confinement have attracted interest from the AdS/QCD community as a means of more closely capturing certain aspects of QCD than the simple hard wall.

This paper is organized as follows. In Section~\ref{sec:eecreview} we review the holographic computation of energy correlators in CFTs with gravitational duals. Then in Section~\ref{sec:eecsingeneralgeometry} we consider a general soft-wall model. Remarkably, we find that the Einstein equations admit exact, linear shockwave solutions, just as in the pure AdS scenario. We use this result to generalize the Hofman--Maldacena calculation to soft-wall confinement. In Section~\ref{sec:results} we demonstrate the calculation with two examples which are theoretically compelling and simple enough to obtain closed-form analytical solutions for the shockwaves. We study a linear dilaton geometry which leads to a gapped continuum spectrum below the confinement scale, and a ``quadratic dilaton'' model that yields linear Regge trajectories. In both cases the resulting two-point energy correlator differs from the hard-wall model. We conclude in Section~\ref{sec:conslusions}, offering comments on the interpretation of our results and discussing their ramifications for the open problems posed by~\cite{csaki2024holographicenergycorrelatorsconfining}.

\section{Holographic energy correlators}
\label{sec:eecreview}

Below we briefly review the definition of energy flow operators and energy correlators, and how to calculate them in a CFT. Physically, energy flow operators can be thought of as idealized calorimeters positioned at various points on the celestial sphere that detect radiation comprised of CFT stuff. This radiation comes from a CFT excitation that is sourced by a perturbation external to the CFT, localized near the origin, and subsequently propagates out to future null infinity. To compute these operators, it is useful to work in lightcone coordinates $x^\pm$ and to perform a conformal transformation mapping future null infinity to $x^+ = 0$. This is described in more detail below. 

We then review the standard holographic computation of energy correlators in pure AdS~\cite{Hofman_2008}, emphasizing those aspects which will require modification for more general holographic models. We consider a strongly-coupled 4D CFT which admits a weakly-coupled gravitational dual on 5D AdS. The energy depositions on the calorimeters mentioned earlier are described by various components of the energy-momentum tensor, which holographically correspond to sourcing bulk gravitons. The $n$-point energy correlator for this process is an $n+2$-point correlator ($n$ insertions of the energy-momentum tensor sourcing gravitons and 2 insertions of the external perturbation) that can in principle be computed in AdS/CFT using Witten diagrams, although they are technically involved. Hofman and Maldecena \cite{Hofman_2008} provided an alternative holographic method for the computation of these correlators. Their prescription involves computing shockwave geometries, which we describe in more detail below. We first review the calculation in pure AdS following~\cite{Hofman_2008,Hofman:2009ug,Belin:2020lsr}, then we summarize how the calculation is modified when an IR brane is included as in \cite{csaki2024holographicenergycorrelatorsconfining}. 

Lastly, we point out how the holographic result for the energy correlators is related to what is typically presented in the collider physics literature (e.g. \cite{CMS-PAS-SMP-22-015, alicecollaboration2024exposingpartonhadrontransitionjets}).

\subsection{Energy correlators and holography}
\label{eecsandholography}

The energy correlator for a localized perturbation sourced by an operator $\mathcal{O}$ is defined as
\begin{align}
    \braket{
    \mathcal{E}(\hat{n}_{1}) ... 
    \mathcal{E}(\hat{n}_{n})
    }
    = 
    \frac{
    \bra{0} \mathcal{O^{\dagger}}
    \mathcal{E}(\hat{n}_{1}) ... 
    \mathcal{E}(\hat{n}_{n})
    \mathcal{O}\ket{0}  } 
    {\bra{0} \mathcal{O}^{\dagger}\mathcal{O}\ket{0}},
\end{align}
where $\mathcal{E}(\hat{n})$ is an energy flow operator, measuring the energy deposited in a calorimeter on the sphere at infinity at a polar angle $\theta$ and azimuthal angle $\phi$ (specified by the unit vector $\hat{n}$). The energy flow operators are defined by the integral
\begin{align}
    \mathcal{E}(\hat{n}) 
    = 
    \lim_{r\rightarrow\infty}
    r^{2}
    \int_{0}^{\infty} dt
    \, T_{0i}(t, r\hat{n}^{i} )\, \hat{n}^{i},
    \label{eq:eflow1}
\end{align}
where the integral over time ensures that $\mathcal{E}$ captures all energy flowing into the calorimeter, the limit $r\rightarrow\infty$ corresponds to the celestial sphere, and the $r^{2}$ accounts for the fact that $T_{0i} \hat{n}^i$ is an energy \textit{flux}. In Eq.~\eqref{eq:eflow1}, we take both $t$ and $r$ separately to infinity, but the order is ambiguous as written. In a CFT, $\mathcal{E}$ will capture all (gapless) excitations if $t+r$ is taken to infinity while $t-r$ is held fixed \cite{Belitsky:2013xxa,Chen:2019bpb}.

When we consider $\mathcal{E}$ for a CFT with a holographic dual, it will prove useful to work in lightcone coordinates ($x^{\pm} = x^{0}\pm x^{3}$). The energy flow operator is then expressed as
\begin{align}
    \mathcal{E}(\hat{n}) 
    = 
    \lim_{x^{+}\rightarrow\infty}
    \frac{(x^{+})^{2}}{4}
    \int_{-\infty}^{\infty} dx^{-} \,
    T_{--}(x^{+}, x^{-}, x^{\perp}),
\end{align}
where $x^{\perp} = x^{1,2}$. Furthermore, for computations it is useful to make a conformal transformation
\begin{align}
    x^{+} \rightarrow -\frac{l^{2}}{x^{+}} , 
    ~~~~~
    x^{-} \rightarrow x^{-} - \frac{|x^{\perp}|^{2}}{x^{+}}, 
    ~~~~~
    x^{\perp} \rightarrow l\frac{x^{\perp}}{x^{+}},
    \label{eq:conftrans}
\end{align}
where $l$ is an arbitrary length scale. The lightcone and spherical coordinates are related via $x^{\pm} = (t-r) + r(1\pm \cos \theta)$, $x^{1} + ix^{2} = r \sin \theta\,e^{i\phi}$. Hence, this conformal transformation in the $r \rightarrow \infty$ limit reduces to 
\begin{align}
    x^{+} \rightarrow 0, 
    ~~~~~
    x^{-} \rightarrow \frac{2(t-r)}{1+\cos\theta}
    ~~~~~
    x^{1}+ix^{2} \rightarrow le^{i\phi}\tan(\theta/2) ,
    \label{eq:ConfTransLimit}
\end{align}
showing that it maps the celestial sphere at future null infinity to the $(x^{1}, x^{2})$ plane at $x^{+}=0$. Light rays on the celestial sphere separated by an angle $\theta$ are mapped to parallel lines on the $x^{+}=0$ plane with a transverse separation $x^{\perp}$. The energy flow operator on this plane takes the simple form (setting $l=1$)
\begin{align}
    \mathcal{E}(\hat{n}) 
    = 
    \Big(1+|x^{\perp}|^{2}\Big)^{3}
    \int_{-\infty}^{\infty}
    dx^{-}T_{--}(x^{+}=0, x^{-}, x^{\perp}).
\end{align}

For a CFT with a holographic dual, the placement of an energy flow operator at a particular transverse coordinate $y^{\perp}$ can be mapped to a deformation of the AdS geometry. To see this more precisely, imagine perturbing the CFT action as
\begin{align}
    \delta S_{\text{CFT}} 
    = 
    \epsilon \int d^4 x T_{--} 
    \delta(x^{+}) \delta^{2}(x^{\perp}-y^{\perp}).
\end{align}
This essentially inserts an exponentiated energy flow operator in the path integral.
Holographically, it sources the $++$ component of the graviton, perturbing the metric:
\begin{align}
    ds^{2} = ds_{\text{AdS}}^{2} + 
    \frac{\epsilon}{z^{2}}\delta(x^{+})
    f(x^{\perp}-y^{\perp},z)(dx^{+})^{2} ,
\end{align}
where
\begin{equation}
    ds_{\text{AdS}}^{2} = (R/z)^{2}(dx^{+}dx^{-}-(dx^{\perp})^{2} - dz^{2}).
\end{equation}
The Einstein equations for a metric of this form reduce to a differential equation for the ``shockwave'' $f(x^{\perp},z)$: 
\begin{align}
    \frac{3}{z} \partial_{z}f -
    (\partial_{1}^{2} + \partial_{2}^{2} + 
    \partial_{z}^{2})f = 0.
    \label{eq:shockwavediffeq}
\end{align}

It is remarkable that the form of the shockwave is given by a linear equation. What is even more remarkable is that it remains linear in more generic warped backgrounds, as we will show in Section~\ref{sec:eecsingeneralgeometry}. The linearity of this equation facilitates the computation of energy correlators, since we can superpose shockwaves to study multiple insertions of energy flow operators at different points $y_{1}^{\perp}$, $y_{2}^{\perp}$, etc.

The boundary condition for the shockwave at $z = 0$ is $f \sim \delta^{2}(x^{\perp})$, since the insertion is localized on the $(x^{1}, x^{2})$ plane. The shockwave in pure AdS is then given by
\begin{align}
    f(x^{\perp}, z) = 
    \frac{z^{4}}{(z^{2} + |x^{\perp}|^{2})^{3}}.
    \label{adsshock}
\end{align}
It is important to be concrete about how this result changes under the conformal transformation mapping the celestial sphere to the null plane. The 4D version of this transformation, Eq.~\eqref{eq:conftrans}, generalizes to AdS by scaling
\begin{equation}\label{eq:conftransAdS}
    z \rightarrow \frac{z}{x^+}, \quad x^- \rightarrow x^- - \frac{\abs{x^\perp}^2 + z^2}{x^+}
\end{equation}
and $x^+, x^\perp$ behave as in Eq.~\eqref{eq:conftrans}. We have set $l = 1$. One can check that this is an isometry of AdS (setting $R = 1$) and that the shockwave transforms by picking up an overall factor of $(x^{+})^{2}$.

Ultimately, the goal is to compute the energy-energy correlator (EEC) in a CFT that is dual to this AdS setup; i.e., there should be a way to relate the shockwave Eq.~\eqref{adsshock} with EECs in the dual CFT. To do so we must specify the external source that perturbs the CFT. For simplicity we shall assume an external scalar source with four-momentum $q^{\mu} = (q, \vec{0})$ 
(which is relevant for e.g. an $e^{+}e^{-}$ collision in the center-of-mass frame with $q = \sqrt{s}$). Holographically, this corresponds to the excitation of a bulk scalar field $\Phi(x^{\mu},z) = \Phi(x^{\mu})\phi(z)$. One can study the equation of motion for the scalar in the presence of two shockwaves, corresponding to the insertion of two exponentiated energy flow operators in the path integral. This leads to 
\begin{align}
    \langle 
    e^{\epsilon_{1}\mathcal{E}(y_{1}^\perp)}
    e^{\epsilon_{2}\mathcal{E}(y_{2}^\perp)}
    \rangle
    \sim 
    \int \frac{dz}{z^{3}} d^{2}x^{\perp} dx^{-}
    i\phi^{*}
    \bigg( 
    e^{-\epsilon_{1}\tilde{f}(y_{1}^{\perp}) \partial_{-}}
    e^{-\epsilon_{2}\tilde{f}(y_{2}^{\perp})  
    \partial_{-}}
    \bigg) 
    \partial_{-}\phi + c.c. , 
    \label{eq:eecinsertion}
\end{align}
where $\tilde{f}(y^{\perp}) = (1 + (y^{\perp})^{2})^{3}f(x^{\perp}-y^{\perp}, z)$. To evaluate the EEC one takes the leading order term $\sim \epsilon_{1}\epsilon_{2}$ and evaluates the integral using the AdS wavefunction $\phi$.

In AdS the wavefunction evaluated at the boundary, $x^+ = 0$, is proportional to a delta function: $\phi \sim e^{iqx^{-}/2}\delta^{2}(x^{\perp})\delta(z-1)$.  For $y_{1}^{\perp} = 0$ and $y_{2}^{\perp} = y^{\perp}$, the leading order term in the above expression integrates to
\begin{align}
    \braket{\mathcal{E}(0)\mathcal{E}(y^{\perp})} 
     \sim \Big(1 + (y^{\perp})^{2}\Big)^{3}f(y^{\perp}, z=1). 
    \label{eq:eecfromshock}
\end{align}
Substituting the form of the AdS shockwave written in Eq.~\eqref{adsshock}, the correlator is a constant: $\braket{\mathcal{E}\mathcal{E}} \sim 1$. This signals a spherically symmetric energy deposition, which is to be expected from a pure strongly-coupled CFT (up to stringy corrections). Note that we are not keeping track of the normalization of the EEC, since in this work we are interested only in the angular dependence.

\subsection{Energy correlators with an IR brane}
Surprisingly, in the presence of an IR brane, the above calculations can still be carried through with some modifications, as outlined in \cite{csaki2024holographicenergycorrelatorsconfining}.  Consider cutting off AdS with an IR brane at $z=z_{\text{IR}}$. This geometry is known to be dual to a strongly-coupled CFT that confines at a scale $\sim 1/z_{\text{IR}}$ and is simply the famous Randall--Sundrum (RS1) model \cite{Randall_1999} with the ultraviolet (UV) brane sent to the AdS boundary. We expect the presence of this new scale to be reflected in the EEC by seeing two different scaling regimes on either side of $z_{\text{IR}}^{-1}$. Above $z_{\text{IR}}^{-1}$ we should see the (constant) CFT expectation for the EEC emerge, and below it a new scaling law should take over.  

The important modification we must make from the full AdS case is to change the boundary condition on the shockwave equation, Eq.~\eqref{eq:shockwavediffeq}. In pure AdS, regularity at $z\rightarrow\infty$ is imposed, but with the addition of an IR brane the space is cut off at $z_{\text{IR}}$, leading to a Neumann boundary condition
\begin{align}
    \partial_{z} f(x^{\perp}, z) 
    \bigg|_{z=z_{\text{IR}}}
    = 
    0. 
\end{align}
By rescaling the shockwave and Fourier transforming over the $x^\perp$ coordinates, the shockwave equation Eq.~\eqref{eq:shockwavediffeq} takes on a simple Schr\"odinger-like form:
\begin{align}
    \left[-\partial_z^2  + k^{2} + \frac{15}{4z^{2}} \right] g(k^\perp, z) = 0, \quad g(k^{\perp}, z) = z^{-3/2}\int e^{-ik^{\perp}\cdot x^{\perp}} f(x^{\perp}, z) dx^{\perp}.
    \label{eq:RSshockwaveEq}
\end{align}
This equation can be solved in terms of Bessel functions, giving the RS1 shockwave
\begin{align}
    f(r, z) = \frac{1}{8}
    \int_{0}^{\infty} dk \, 
    J_{0}(kr)k^{3}z^{2} 
    \bigg[ 
    K_{2}(kz) + 
    \frac{K_{1}(kz_{\text{IR}})}{I_{1}(kz_{\text{IR}})}
    I_{2}(kz) 
    \bigg]
    \label{eq:RSshockwave}
\end{align}
where $r \equiv |x^{\perp}|$ (different from the $r$ in Section \ref{eecsandholography}). The normalization is chosen such that $\lim_{z\rightarrow 0} z^{3/2} g(k^\perp, z) = 1/4$, which ensures that in the $z_{\text{IR}}\rightarrow\infty$ (pure AdS) limit the EEC is 1.

The conformal transformation of Eq.~\eqref{eq:conftransAdS} is no longer an isometry once one introduces the IR brane. Since the IR brane corresponds to a spontaneous breaking of conformal symmetry, the IR brane location should transform covariantly, i.e. we should take $\zir \rightarrow \zir / x^+$. Then the transformation is an isometry of the metric and the shockwave Eq.~\eqref{eq:RSshockwave} again transforms with an overall factor of $(x^{+})^{2}$. After performing the transformation the brane is located at $z = x^+ \zir$.

The expression for the EEC then follows a similar procedure to the previous subsection; it amounts to computing the integral in Eq.~\eqref{eq:eecinsertion}.
In the limit where the energy of a scalar source is large compared to the confinement scale, $q \zir \gg 1$, one recovers the AdS wavefunction, exhibiting the same delta-function behavior in the $x^+ \rightarrow 0$ limit. After careful treatment of the $x^+ \rightarrow 0$ limit, one can perform the integration in Eq.~\eqref{eq:eecinsertion} and expand to linear order in $\epsilon_{1,2}$, yielding the same form found in Eq.~\eqref{eq:eecfromshock}. The only difference is that in RS1 the shockwave is a more complicated expression, Eq.~\eqref{eq:RSshockwave}, than the simpler AdS case, Eq.~\eqref{adsshock}. One then finds the correlator is \textit{not} constant; we will show this result in Section~\ref{sec:results} when comparing the correlators among other geometries.

\subsection{CFT and collider conventions}\label{conventions}

To wrap up the review of holographic energy correlators we clarify the different conventions in the literature. The EEC written in Eq.~\eqref{eq:eecfromshock} is, as typically defined in perturbative QCD (for example \cite{Dixon_2018}), a differential cross section in the angular separation $\chi$ weighted by the energies of final state particle pairs, $E_{i}$ and $E_{j}$, with angular separation $\chi_{ij}$:
\begin{align}
    \frac{d\sigma_{E}}{d\text{cos}\,\chi} 
    = 
    \sum_{i,j}
    \int
    E_{i}E_{j}
    \delta(\text{cos}\,\chi_{ij}
    - 
    \text{cos}\,\chi) 
    \frac{d\sigma}{\sigma},
\end{align}
It is normalized as
\begin{align}
    \int_{-1}^{1}d\text{cos}\,\chi\,
    \frac{d\sigma_{E}}{d\text{cos}\,\chi} 
    = q^{2},
\end{align}
where $q^{\mu}$ is the total 4-momentum (the same $q$ as in our example with a scalar source). This can be checked by considering the properly normalized, holographic, one-point energy correlator in pure AdS, which is given by $\braket{\mathcal{E} } = q/4\pi$~\cite{Hofman_2008}. Integrating over $d\Omega = d\text{cos}\,\theta \, d\phi$ gives back the total energy $q$ as expected. We can therefore write 
\begin{align}
    \braket{\mathcal{E}\mathcal{E}} \sim \frac{d\sigma_{E}}{d\text{cos}\,{\theta}},
\end{align}
where $\theta = \theta_{1}-\theta_{2}$ is the difference in polar angles between the two detectors on the celestial sphere. We also ignore the azimuthal piece since all correlators we compute have axisymmetry.

In Section \ref{sec:results} our results for the EEC are presented in two conventions. In the first convention we plot $\frac{d\sigma_{E}}{d\text{cos}\,\theta}$ against $r$, where $r=r_{1}-r_{2}$ is simply $\tan\frac{\theta_{1}}{2}$ assuming we take $\theta_{2} = 0$. We refer to this as the ``CFT convention'' since it is more directly derived from the CFT literature \cite{Hofman_2008}.  

In the second convention, we show our results in terms of pseudorapidity $\eta=\eta_{1}-\eta_{2}$. The plot axes in this convention are $\eta\frac{d\sigma_{E}}{d\eta}$ and $\eta$, where, choosing $\theta_{2} = \pi/2$: 
\begin{align}\label{eq:pseudorapidity}
    \eta  = 
    -\text{log}
\,\tan
\bigg( 
\frac{\theta}{2} + \frac{\pi}{4}
\bigg).
\end{align}
Note that this choice of $\theta_{2}$ changes the relation between $r$ and the polar angles on the sphere: 
\begin{align}
    \theta = 2
    \bigg(
    \text{arctan}(r+1) - \frac{\pi}{4} 
    \bigg). 
\end{align}
This convention is similar to energy correlators found in collider physics literature (i.e. \cite{CMS-PAS-SMP-22-015, alicecollaboration2024exposingpartonhadrontransitionjets}), which are typically differential distributions in the angular distance $\Delta R$. Since our correlators all have axisymmetry, we take $\Delta R \sim \eta$. 

For completeness we write the relation between the two conventions:
\begin{align}
    \eta\frac{d\sigma}{d\eta} 
    =
    \text{log}
\,\tan
\bigg( 
\frac{\theta}{2} + \frac{\pi}{4}
\bigg) 
    \sin\theta\cos\theta
    \frac{d\sigma}{d\cos\theta}.
    \label{eq:jacobian}
\end{align}

\section{Energy correlators in a general warped geometry}
\label{sec:eecsingeneralgeometry}

We will now generalize the holographic calculation of energy correlators to an arbitrary warped geometry
\begin{equation}\label{eq:arbitrarymetric}
    ds^2 = e^{-2A(z)} \left( dx^+ dx^- - \abs{ dx^\perp }^2 - dz^2 \right)
\end{equation}
where $A(z)$ is the warp factor. 

In practice, we are interested in deforming the geometry away from AdS in the IR to model effects of confinement. Thus, we expect the warp factor to include a dimensionful scale corresponding to the scale of confinement. We therefore consider a warp factor
\begin{equation}\label{eq:arbitrarywarpfactor}
    A(z) = \log \frac{z}{R} + w(z/\zir) ,
\end{equation}
where $1/\zir$ is the confinement scale. The function $w$ characterizes the deviation from AdS. We will assume that $w(z/\zir)$ approaches $0$ as $z \rightarrow 0$ such that the metric is asymptotically AdS in this limit. The dual interpretation of this is that the CFT is deformed by relevant operators which are negligible at energies well above the confinement scale.

To solve the Einstein equations for this geometry a bulk scalar field is needed, in contrast to AdS where a bulk cosmological constant suffices. Hence we introduce a real scalar $\phi$, whose action is
\begin{equation}
    \int d^5 x \sqrt{g} \left[ \frac{1}{2} \left( \nabla \phi \right)^2 - V(\phi) \right]
\end{equation}
where $V(\phi)$ is the scalar potential. We assume that the scalar acquires a $z$-dependent vev $\phi(z)$.

The coupled Einstein-scalar equations of motion are given by
\begin{equation}\label{eq:warpfactoreom}\begin{split}
    \left( \phi'(z) \right)^2 &= \frac{3}{\kappa^2} \left( A'(z)^2 + A''(z) \right) , \\
    V(\phi) &= -\frac{6}{\kappa^2} e^{2A(z)} A'(z)^2 .
\end{split}\end{equation}
Here $\kappa$ is related to the 5D gravitational constant $G_5$ by $\kappa^2 = 8 \pi G_5$.

\textit{A priori} it is not clear whether it is even possible to generalize the holographic energy correlator calculation. The purpose of this section is to establish this. We need to show three things:
\begin{itemize}
    \item The conformal transformation used to map the celestial sphere to the null plane remains a symmetry of the geometry in Eq.~\eqref{eq:arbitrarymetric}.
    \item Linear shockwave solutions to the Einstein equations exist for any warp factor.
    \item The wavefunction for a scalar source is identical to the AdS case in the high-energy limit.
\end{itemize}
This is sufficient for the RS energy correlator calculation to carry over to an arbitrary warp factor.

The first of these conditions is easy to verify. We note the transformation of Eq.~\eqref{eq:conftransAdS} is an isometry of the metric so long as we also take $\zir \rightarrow \zir / x^+$. This is analogous to the RS case, where the location of the IR brane must be rescaled as $\zir \rightarrow \zir / x^+$ for Eq.~\eqref{eq:conftransAdS} to be an isometry. One can interpret this procedure as promoting the confinement scale to a spurion which tranforms covariantly under the conformal group, thus restoring the conformal symmetry.

\subsection{Shockwaves}

Next we consider shockwave perturbations about the background metric. We modify the metric to $ds^2 + \delta ds^2$, where
\begin{equation}
    \delta ds^2 = \epsilon e^{-2A(z)} \delta(x^+) f(x^\perp, z) \left(dx^+ \right)^2 .
\end{equation}
This modifies the Einstein tensor $G$ and the stress-energy tensor $T$ as follows:
\begin{equation}\begin{split}
    \delta G &=  \frac{\epsilon}{2} \delta(x^+) \left( dx^+ \right)^2 \left[ -6 \left( A'(z)^2 - A''(z) \right) -3 A'(z) \partial_z + \partial_z^2  + \partial_1^2 + \partial_2^2 \right] f(x^\perp, z) , \\
    \delta T &=  \epsilon \delta(x^+) \left( dx^+ \right)^2 \left[ e^{-2A(z)} V(\phi) + \phi'(z)^2 \right] .
\end{split}\end{equation}
These equations are exact, not perturbative expansions in $\epsilon$.

Using the equations of motion for the background, Eq.~\eqref{eq:warpfactoreom}, we obtain a simple linear equation of motion for $f$,
\begin{equation}\label{eq:shockwaveeomgeneral}
    \left( 3 A'(z) \partial_z - \partial_z^2 -  \partial_1^2 - \partial_2^2 \right) f(x^\perp, z) = 0 .
\end{equation}
It is remarkable that the shockwaves are linear even in an arbitrary warped background. This means that shockwaves about any background can be superposed to compute energy correlators, just like AdS shockwaves.

We did not need to include a perturbation of the stabilizing scalar $\phi$ in addition to the graviton perturbation to find a solution. This indicates that the shockwaves do not mix with Kaluza--Klein (KK) modes (nor the zero mode) of the scalar. More explicitly, we can consider fluctuations of the scalar field, $\phi \rightarrow \phi(z) + \varphi(x^\mu, z)$. The equation of motion at linear order in the fluctuation is
\begin{equation}
    0 = 3 A' \varphi' - \varphi'' - \partial_\perp^2 \varphi + 4 \partial_+ \partial_- \varphi - 4 f(x^\perp, z) \delta(x^+) \partial_-^2 \varphi + e^{-2A} \frac{d V}{d\phi} \varphi ,
\end{equation}
where the derivative $dV/d\phi$ is evaluated on the vev of $\phi$. If the shockwave mixed with scalar fluctuations, then there would be a source term proportional to $f$ in this equation. Instead the only dependence on the shockwave is contained in the $f \partial_-^2 \varphi$ term, indicating a three-point, shockwave-scalar-scalar coupling.

\subsection{Wavefunction}

The last piece of the holographic energy correlator calculation is the external source which excites the CFT. Here we will consider the simplest case of a scalar external source with energy $q$ and no 3-momentum. In the holographic picture this corresponds to studying a bulk scalar field, which we will call $\chi$ to distinguish it from the stabilizing field $\phi$, with the boundary condition $\chi = e^{iqt}$ on the UV boundary at $z = 0$. We are working in the original Poincar\'e coordinates, and we will later perform the conformal transformation given in Eq.~\eqref{eq:conftransAdS}.

Intuitively, we expect that for $q$ much larger than the confinement scale $1/\zir$, the wavefunction should be insensitive to the confining dynamics. That is, in the $q\zir \gg 1$ limit we should recover the AdS wavefunction for the scalar. This is indeed the case in the hard-wall model~\cite{csaki2024holographicenergycorrelatorsconfining}.

We can see this explicitly by studying the scalar equation of motion:
\begin{equation}
    \left( 3 A'(z) \partial_z - \partial_z^2 + \partial^\mu \partial_\mu \right) \chi(x^\mu, z) = 0 .
\end{equation}
We can put this in a Schr\"odinger-like form by rewriting the field as 
\begin{equation}
    \chi = e^{iqt} q^{3/2} e^{3A(z)/2} \psi(z).
\end{equation}
This leads to
\begin{equation}\label{eq:scalareom}
    \left[-\partial_z^2 + V(z) \right] \psi(z) = 0 ,
\end{equation}
where the potential is
\begin{equation}\begin{split}
    V(z) &= \frac{9}{4} A'(z)^2 - \frac{3}{2} A''(z) - q^2 \\
    &= \frac{15}{4z^2} - q^2 + \frac{9 w'(z/\zir)}{2 z \zir} + \frac{9 w'(z/\zir)^2}{4 \zir^2} - \frac{3 w''(z/\zir)}{2 \zir^2} .
\end{split}\end{equation}
The first two terms are what one obtains for pure AdS, while the remaining terms are corrections to the potential of order $z/\zir$.

Consider the pure AdS case, when $w = 0$. In the regime $qz \gg 1$ we have the approximate solution $\psi \sim e^{-iqz}$, with the sign fixed by an outgoing wave boundary condition. Hence the wavefunction is
\begin{equation}\label{eq:adswavefunction}
    \chi_{\rm AdS} \sim e^{iq(t - z)} (qz)^{3/2} .
\end{equation}
In fact one can just solve the equation of motion exactly in terms of Bessel functions, and after applying an outgoing wave boundary condition one finds $\chi_{\rm AdS} = e^{iqt} (qz)^2 J_2(q z)$ (up to a power of $q$ fixed by the scaling dimension, which is related to the bulk mass of the field). Then we can take the large $q$ limit by using the asymptotic expansion $J_2(x) \sim e^{-ix}/\sqrt{x}$ (up to an overall constant and phase). This confirms the form of Eq.~\eqref{eq:adswavefunction}.

When one performs the transformation in Eq.~\eqref{eq:conftransAdS} and goes to the future null boundary of Minkowski space (located at $x^+ = 0$ after the transformation), the scalar wavefunction is delta-function localized,
\begin{equation}
   \chi_{\rm AdS,~after~transformation}(x^+ = 0, x^-, x^\perp, z) \sim \delta(z-1)\delta^2(x^\perp) e^{iqx^-/2}. 
\end{equation}
This is not obvious when one writes the wavefunction in Poincar\'e coordinates as above, but it is easy to see by embedding AdS in Euclidean space, see~\cite{csaki2024holographicenergycorrelatorsconfining,Hofman_2008}. Using the embedding coordinates one can show the wavefunction approaches a Gaussian  for small $x^+$, whose width is given by $x^+ / q z$. As one approaches future null infinity the width of the Gaussian goes to zero, becoming a delta function centered at $x^\perp = 0$ and $z = 1$.

Now consider an arbitrary warp factor in Eq.~\eqref{eq:scalareom}. In the regime $qz, q\zir \gg 1$ this is approximately solved by $\psi = e^{-iqz}$, leading to the asymptotic behavior
\begin{equation}\label{eq:wavefunction}
    \chi \sim e^{iq(t - z)} (qz)^{3/2} e^{3 w(z/\zir)/2} .
\end{equation}
We continue to ignore an overall power of $q$, which is determined by the scaling dimension of the field and does not affect the angular dependence of energy correlators.

We see that the wavefunction in Eq.~\eqref{eq:wavefunction} matches the AdS wavefunction, Eq.~\eqref{eq:adswavefunction}, up to the overall $e^{3w(z/\zir)/2}$ factor. Note that the conformal transformation mapping the celestial sphere to the null plane at $x^+ = 0$ does not change the ratio $z/\zir$.
Thus, after performing the transformation, the wavefunction should match the AdS result (essentially a Gaussian with width $x^+ / qz$), times a factor of $e^{3w(z/\zir)/2}$. Then the wavefunction at small $x^+$ should match the AdS result,
\begin{equation}
    \chi_{\rm after~transformation}(x^+ = 0, x^-, x^\perp, z) \sim \delta(z-1) \delta^2(x^\perp) e^{iqx^-/2} ,
\end{equation}
up to a factor of $e^{3w(1/\zir)/2}$. This factor only affects the overall normalization of the wavefunction, which is unimportant for our purposes.

For this argument to work it is essential that $q \zir \gg 1$. Otherwise, there is no regime where the scalar wavefunction admits an approximate plane wave solution as in Eq.~\eqref{eq:wavefunction}. This matches our intuition that the effects of confinement should not impact a source with energy well above the confinement scale.
 
The upshot is that for a high-energy source, the wavefunction is delta-function localized as in the AdS case. Thus the two-point energy correlator essentially probes the shockwave at $z = 1$, just like in Eq.~\eqref{eq:eecfromshock}. 

\section{Analytical results for specific warp factors}\label{sec:results}

We now focus on two simple confining holographic models where one can actually obtain an analytical expression for the two-point energy correlator. We first study a linear dilaton model, where the warp factor is
\begin{equation}\label{eq:metriclineardilaton}
    A(z) = \log z + \mu z  \quad (\rm{linear~dilaton~/~gapped~continuum}).
\end{equation}
We have set the AdS curvature $R$ to $1$, and $\mu$ corresponds to the scale of confinement, which we will later identify with $1/z_{\text{IR}}$ for comparison with the hard-wall model. This warp factor is interesting from a theoretical perspective because it leads to a gapped continuum spectrum for the KK modes of bulk fields, in contrast to the usual discrete spectrum. It is reasonable to expect that such a drastic modification of the IR spectrum would leave an imprint on energy correlator observables. It is worth noting that gapped continuum spectra have also attracted phenomenological interest in the context of dark matter~\cite{Csaki:2021gfm,Csaki:2021xpy,Csaki:2022lnq,Ferrante:2023fpx,Katz:2015zba,Chaffey:2021tmj}, the hierarchy problem~\cite{Falkowski:2008fz,Bellazzini:2015cgj,Csaki:2018kxb,Cabrer:2009we,Megias:2019vdb,Megias:2021mgj}, supersymmetric models~\cite{Cai:2009ax,Cai:2011ww,Gao:2019gfw}, and the cosmological collider~\cite{Aoki:2023tjm,Hubisz:2024xnj}.

The second model we study involves a warp factor which grows like $z^2$ in the deep IR,
\begin{equation}\label{eq:metricquadraticdilaton}
    A(z) = \log z + \frac{1}{2} \mu^2 z^2  \quad (\rm{quadratic~dilaton~/~linear~confinement}),
\end{equation}
where again $\mu$ represents the confinement scale. One could call this a ``quadratic dilaton'' model in analogy with the linear dilaton. This model leads to a linearly growing mass-squared spectrum for the KK modes, $m_n^2 \sim n$, as opposed to the $m_n^2 \sim n^2$ scaling found in the RS1 scenario~\cite{Karch:2006pv}. Since the linear confinement is reminiscent of Regge trajectories in QCD, geometries of this sort are interesting from the perspective of AdS/QCD.

We can put the shockwave equation of motion, Eq.~\eqref{eq:shockwaveeomgeneral}, into a Schr\"odinger-like form. This is analogous to the manipulations we performed to obtain the scalar wavefunction in Eq.~\eqref{eq:scalareom}. Defining
\begin{align}
    g(z, k^{\perp})  = 
    N
    e^{-3A(z)/2}\int e^{i\vec{k}^{\perp}\cdot \vec{x}^{\perp}}
    f(z, x^{\perp})
    dx^{\perp},
\end{align}
where $N$ is an overall normalization, Eq.~\eqref{eq:shockwaveeomgeneral} can be rewritten as
\begin{align}\label{eq:shockwaveschrodingerform}
    \left[-\partial_z^2 + k^{2} + V(z)\right]g(z,k^\perp) = 0
    , \quad
    V(z) = \frac{9A'(z)^{2}}{4}
    - \frac{3A''(z)}{2},
\end{align}
with $k\equiv |k^{\perp}|$. Since we are only interested in the angular dependence of energy correlators, not their overall magnitude, we choose to normalize our shockwaves by requiring $\lim_{z\rightarrow 0} e^{3A(z)/2} g(z, k^\perp) = 1/4 $. This choice of normalization gives $\langle \mathcal{E} \mathcal{E} \rangle = 1$ for the pure AdS case.

\subsection{Gapped continuum}

The Schrödinger-like potential for the linear dilaton metric, given by Eq.~\eqref{eq:metriclineardilaton}, is 
\begin{align}
    V(z) = 
    \frac{15}{4z^{2}} + 
    \frac{9\mu}{2z} + 
    \frac{9\mu^{2}}{4} .
\end{align}
The solution to Eq.~\eqref{eq:shockwaveschrodingerform} with this potential is a linear combination of the Whittaker $M$ and $W$ functions. Imposing regularity deep in the IR (large $z$) eliminates the $M$ Whittaker function and the UV boundary condition, $g(z) \sim e^{-3A(z)/2} / 4$ as $z\rightarrow 0$, fixes the coefficient of the $W$ function. The result is 
\begin{align}
    g(k,z)
    = 
    \frac{1}{4}
    \frac{3k^{2}(k^{2}-18\mu^{2})}
    {2\beta^{5/2}}
    \Gamma
    \big(-\alpha-\tfrac{3}{2}\big)
    W_{\alpha,2}(\beta z),
\end{align}
where 
\begin{equation}
    \alpha = \frac{-9\mu}{2\beta}, \quad \beta = \sqrt{4k^{2} + 9\mu^{2}}.
\end{equation}
Fourier transforming back to position space in $x^\perp$, the shockwave is given by
\begin{align}
    f(x^{\perp},z) = 
    \int_{0}^{\infty}
    J_{0}(kr) k z^{3/2} e^{\frac{3}{2}\mu z} 
    g(k,z).
\label{eq:LDshockwave}
\end{align}

\subsection{Linear confinement}

The Schrödinger-like potential for the quadratic dilaton metric, given by Eq.~\eqref{eq:metricquadraticdilaton}, is
\begin{align}
    V(z) = 
    \frac{15}{4z^{2}} + 
    \frac{6\mu^{2}}{2} + 
    \frac{9\mu^{4}z^{2}}{4}.
\end{align}
The solution can be expressed as a linear combination of a Tricomi confluent hypergeometric function $U$ and a generalized Laguerre polynomial $L$. Similarly to the linear dilaton case, regularity in the IR eliminates one of the two solutions and the UV boundary condition fixes the overall coefficient. The solution is
\begin{align}
    g(k,z) = 
    \frac{1}{8}
    \frac{k^{2}(k^{2} + 6\mu^{2})}{18 \mu^{4}}
    \Gamma
    \big(\tfrac{k^{2}}{6\mu^{2}}\big)
    z^{-3/2}e^{-\frac{3}{4}\mu^{2}z^{2}}
    U_{\frac{k^{2}}{6\mu^{2}},-1}(\tfrac{3}{2}\mu^{2}z^{2}).
\end{align}
Again, Fourier transforming back to position space gives the form of the shockwave 
\begin{align}
    f(x^{\perp},z) = 
    \int_{0}^{\infty} 
    J_{0}(kr)
    k z^{3/2} 
    e^{\frac{3}{4}\mu^{2}z^{2}}
    g(k,z).
\label{eq:QDshockwave}
\end{align}

\subsection{Results}

Now we use the shockwaves derived in Eqs.~\eqref{eq:LDshockwave} and \eqref{eq:QDshockwave} to compute energy correlators. Recall that in the limit of a high-energy scalar source with no 3-momentum, the two-point energy correlator can be calculated from the shockwave using Eq.~\eqref{eq:eecfromshock}:
\begin{align}
    \braket{\mathcal{E}(0)\mathcal{E}(r)} 
    \sim 
    \Big(1+r^{2}\Big)^{3}
    f(r, z=1) ,
\end{align}
where $r = \abs{x^\perp}$.

\begin{figure}
\centering
\includegraphics[width=6in]{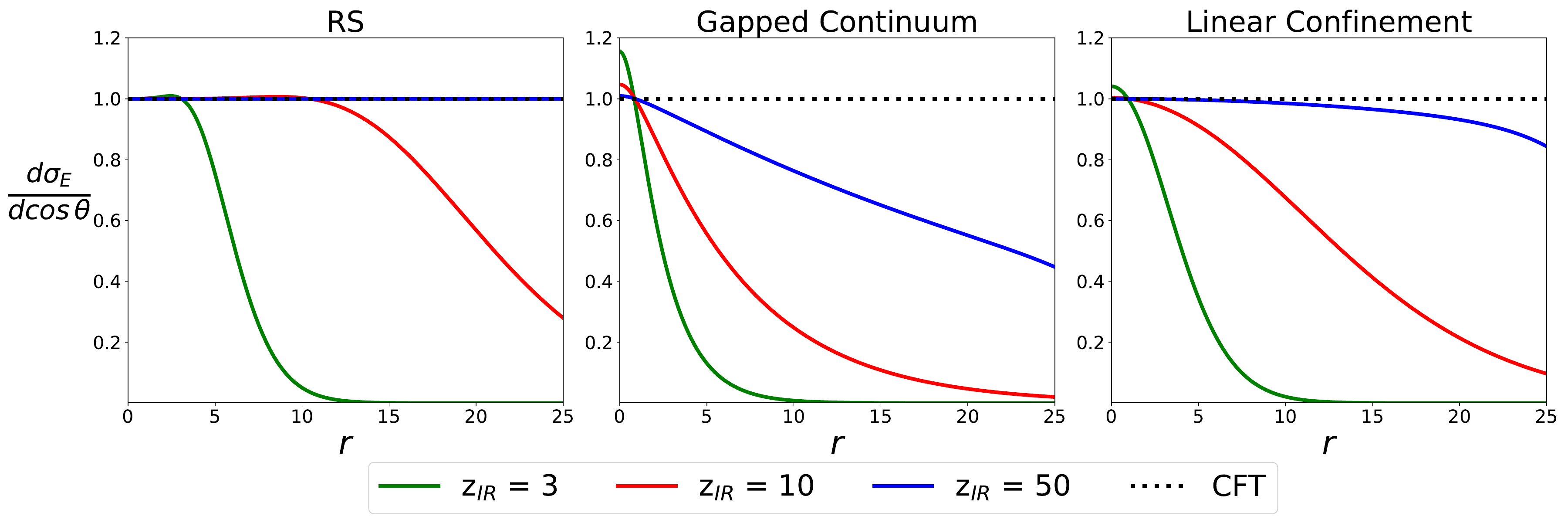}
\caption{The holographic two-point energy correlator in the minimal hard-wall~\textit{(left)}, linear dilaton~\textit{(center)}, and quadratic dilaton~\textit{(right)} models. For each model we show three choices of the IR scale $\mu^{-1} = \zir$: $3$~\textit{(green)}, $10$~\textit{(red)}, and $50$~\textit{(blue)}. The dotted black line is the constant correlator obtained in AdS. We plot $d \sigma_E / d \cos \theta$ as a function of separation on the null plane $r$, corresponding to the CFT convention of Section~\ref{conventions}; the overall normalization is arbitrary.}
\label{fig:EECs_CFT}
\end{figure}

\begin{figure}
\centering
\includegraphics[width=6in]{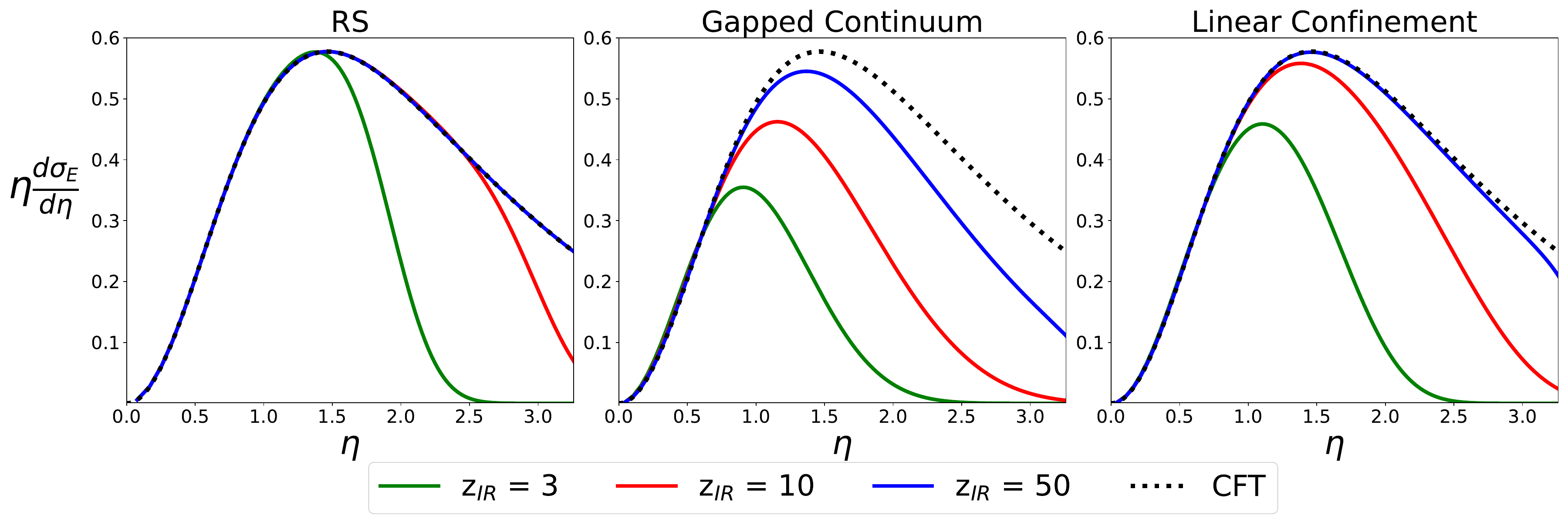}
\caption{Same as Fig.~\ref{fig:EECs_CFT}, but with the results plotted in the collider convention defined in Eqs.~\eqref{eq:pseudorapidity} and~\eqref{eq:jacobian}, as opposed to the CFT convention.}
\label{fig:EECs_Coll}
\end{figure}

In Fig.~\ref{fig:EECs_CFT} we plot our main result: the two-point correlator for the linear dilaton (center panel) and quadratic dilaton (right panel) models as a function of separation on the null plane $r$, working in the CFT convention explained in Section~\ref{conventions}. For comparison, the left panel of Fig.~\ref{fig:EECs_CFT} shows the two-point correlator for the minimal hard-wall model, first computed in ~\cite{csaki2024holographicenergycorrelatorsconfining}. We also include a comparison to the trivial strongly-coupled CFT result, which is just a constant. We present results for three choices of the IR scale, $\mu = 1/3$ (green), $\mu = 1/10$ (red), and $\mu = 1/50$ (blue); for comparing to the hard-wall model we take the brane location $\zir = 1/\mu$.
In Fig.~\ref{fig:EECs_Coll} we present the same data in the collider convention defined by Eqs.~\eqref{eq:pseudorapidity} and~\eqref{eq:jacobian}. 

In all cases, the correlator falls off at large $r$. This differs from the constant result one obtains for a pure strongly-coupled CFT, and it suggests the presence of confinement. We will later confirm that the decay is exponential. At smaller $r$, we see that the correlators for the gapped continuum and linear confinement geometries differ markedly from each other, as well as from the RS model. This means that, at least in principle, one could distinguish between these models by measuring the two-point correlator.

Let us compare the hard wall (left panel) and linear confinement (right panel) correlators in more detail. In the former, one has a nearly constant correlator (matching the strongly-coupled CFT expectation) up to $r \sim \zir$, followed by an exponential decay. For $\zir \gg 1$, this implies that when one maps back to the celestial sphere, the two-point correlator differs from pure AdS case only in the back-to-back limit. Indeed this is the behavior observed in Fig.~\ref{fig:EECs_Coll}. In the linear confinement model one has a more gradual transition between the constant and decaying regimes, and the deviation from the constant correlator starts to occur at somewhat smaller $r$. Thus one can distinguish this model from the hard wall by measuring the EEC in the back-to-back regime.

The gapped continuum (middle panel) case differs strikingly from the other models. From Fig.~\ref{fig:EECs_CFT}, we see an exponential decay of the correlator even for small separation on the null plane. Comparing the gapped continuum correlator at different $z_{\text{IR}}$, in all cases it appears that the decay sets in around $r \sim 2$. This suggests that the quick transition to an exponential decay occurs irrespective of the confinement scale.
It is also interesting to recognize the way in which the gapped continuum correlator approaches the strongly-coupled CFT result in the large $z_{\text{IR}}$ limit, and how it is distinct from the other two cases. In the hard-wall and linear confinement models, the exponential decay starts at larger values of $r$ for larger values of $z_{\text{IR}}$, and eventually for $z_{\text{IR}}\rightarrow\infty$, the entire correlator is flat and it never starts falling off. In the gapped continuum model, however, the exponential decay always starts near $r\sim 2$, but we see that the decay constant decreases for larger $z_{\text{IR}}$. Eventually the decay constant approaches 0, so the correlator approaches the expected CFT result.  This seems to indicate that the confining dynamics in the gapped continuum model is qualitatively distinct from that of the other two models.

\begin{figure}
\centering
\includegraphics[width=6in]{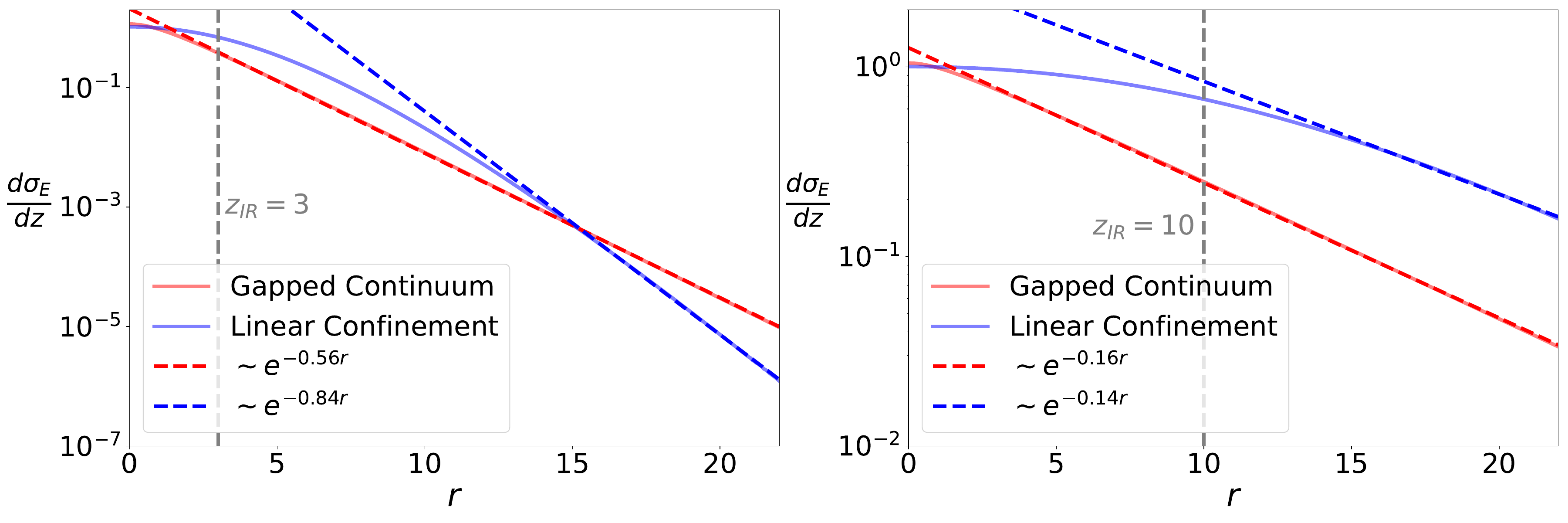}
\caption{Exponential fits for the previously computed EECs for the linear dilaton~\textit{(red)} and quadratic dilaton~\textit{(blue)} scenarios. The left panel is for $z_{\text{IR}}=3$ and the right panel is for $z_{\text{IR}}=10$. }
\label{fig:LDandQDfits}
\end{figure}

In \cite{csaki2024holographicenergycorrelatorsconfining} it was argued analytically that the correlator must decay faster than a power law, and numerically it was found that it decays exponentially for large $r$. Here there are fewer analytical handles to study the asymptotics of the integrals Eq.~\eqref{eq:LDshockwave} and Eq.~\eqref{eq:QDshockwave}, but it can still be seen numerically that one has an exponential decay. In Fig.~\ref{fig:LDandQDfits} we fit the linear dilaton and quadratic dilaton correlators to an exponential function, showing that for large $r$, both fall off like $\sim e^{-\beta r}$. 

For the gapped continuum, the fit yields the decay constant $\beta = 0.56$ and $0.16$ for $z_{\text{IR}} = 3$ and $10$ respectively. As mentioned above, the exponential decay for this case turns on almost immediately, at $r\sim 2$. The fit shown includes points only above $r=1.5$. For the linear confinement case, the fit yields $\beta = 0.84$ and 0.14 for $z_{\text{IR}} = 3$ and 10 respectively, and in contrast the exponential decay doesn't appear until larger values of $r$ ($\sim 15)$. In this case only points above $r=15$ are included in the fit.

Note that there is a wide body of literature on the asymptotic behavior of Hankel transforms~\cite{MR493193,MR804942,MR811192,MR903988,MR1851050,MR3149389}. It would be interesting to attempt to derive the asymptotic behavior we observe in Fig.~\ref{fig:LDandQDfits} using some of these analytical techniques. We have not attempted this, but we have a simple argument that the two-point correlator must decay faster than a power law for any choice of the warp factor. The equation of motion for the shockwave, after performing a Fourier transform over $x^\perp$, depends on the momentum $k^\perp$ only through its magnitude squared, $k^2$. This is manifest in Eq.~\eqref{eq:shockwaveschrodingerform}. Therefore, the Taylor series expansion of the shockwave $f(k^\perp, z)$%
\footnote{We hope the notation is obvious, but to be clear the shockwave in Fourier space is
\begin{equation}
    f(k^{\perp},z) = \int e^{i\vec{k}^{\perp}\cdot \vec{x}^{\perp}} f( x^{\perp}, z) d^2 x^{\perp}.
\end{equation}}
 only contains even powers of $k$ (it can also contain terms with logarithms like $k^{2n} \log k$, but this does not affect our argument). Then the theorems due to Wong in~\cite{MR493193} imply $f(x^\perp, z)$ approaches its limiting value faster than any power law.

\section{Conclusions}
\label{sec:conslusions}

In this paper we have demonstrated how to generalize the shockwave method for computing holographic energy correlators to general warped backgrounds. This was possible because linear shockwave solutions exist about a geometry with an arbitrary warp factor. To our knowledge this class of solutions to the Einstein equations has not been previously appreciated in the literature.

We then calculated the two-point energy correlator in two simple soft-wall models of confinement where we could obtain analytical solutions. Importantly, the correlators in different models of confinement differed from each other. This suggests that the dynamics of confinement are imprinted upon energy correlators.

We remark that there were two features common to all the cases we studied, including the hard-wall model~\cite{csaki2024holographicenergycorrelatorsconfining}. We found little deviation from the full AdS (no confinement) computation in the collinear limit and the correlator asymptotically exhibited an exponential decay in $x^\perp$ in the back-to-back limit. It is important to emphasize that this appears different from how confinement manifests in QCD energy correlators, which hints at fundamental differences between confinement in QCD and in holographic models. This is perhaps not surprising since it is well-known that holographically modeling jets requires the incorporation of stringy dynamics in the bulk~\cite{Csaki:2008dt}.

Our results suggest that these common features could be general aspects of holographic confinement.
We argued in Section~\ref{sec:results} that this is indeed the case for the exponential decay, using theorems regarding the asymptotics of Hankel transforms. The collinear behavior is peculiar because one generically expects nonperturbative effects in both the collinear and back-to-back limits. To better understand the collinear limit, it may be important to include the effects of jets. It might also be useful to study corrections to the extreme high-energy limit ($q \zir \gg 1$) taken throughout this paper as well as~\cite{csaki2024holographicenergycorrelatorsconfining}.

On the other hand, it is worth noting that the falloff of the correlator in the back-to-back limit is still in rough qualitative agreement with the QCD EEC. In QCD, however, the falloff in the back-to-back limit arises in perturbation theory at the leading-log due to Sudakov suppression \cite{de_Florian_2005} --- regardless of confinement --- and it receives nonperturbative corrections related to hadronization. This is different than what we see in these holographic models, where the falloff only occurs once we introduce confinement. It is curious that in either case one performs a Hankel transform, resulting from a 2D Fourier transform over transverse coordinates (involving the shockwave in holography and a Sudakov factor in QCD). Given the similar mathematical form despite the different underlying physics, one might further speculate that the EECs in this limit are related. 

This work represents another step in forging a closer connection between holography and energy correlators in confining theories, building upon the initial results of~\cite{csaki2024holographicenergycorrelatorsconfining}. We have shown that changing the IR dynamics of confinement via modifying the 5D geometry at large $z$ manifests observable effects in the two-point energy correlator.

The next natural step is to consider a UV deformation of theory; we focused on the case of an asymptotically AdS metric, so the UV behavior of the dual theory is always the same. Of particular relevance to the real world would be a 5D model which models asymptotic freedom such as~\cite{Csaki:2006ji}, which we plan to study in future work~\cite{futurework}. Some other directions for further inquiry include studying corrections to the high-energy limit and incorporating jets (as mentioned above), as well as considering sources other than scalars.

\acknowledgments
CC and SF are supported in part by the NSF grant PHY-2309456.  CC is supported in part by the US-Israeli BSF grant 2016153. AI is supported by a Mafalda and Reinhard Oehme Postdoctoral Research Fellowship from the Enrico Fermi Institute at the University of Chicago. 

\bibliographystyle{utphys}
\bibliography{ref}

\providecommand{\href}[2]{#2}\begingroup\raggedright\begin{thebibliography}{10}

\bibitem{Hofman_2008}
D.~M. Hofman and J.~Maldacena, ``{Conformal collider physics: Energy and charge
  correlations},'' \href{http://dx.doi.org/10.1088/1126-6708/2008/05/012}{{\em
  JHEP} {\bfseries 05} (2008) 012},
  \href{http://arxiv.org/abs/0803.1467}{{\ttfamily arXiv:0803.1467 [hep-th]}}.

\bibitem{Komiske:2022enw}
P.~T. Komiske, I.~Moult, J.~Thaler, and H.~X. Zhu, ``{Analyzing N-Point Energy
  Correlators inside Jets with CMS Open Data},''
  \href{http://dx.doi.org/10.1103/PhysRevLett.130.051901}{{\em Phys. Rev.
  Lett.} {\bfseries 130} no.~5, (2023) 051901},
  \href{http://arxiv.org/abs/2201.07800}{{\ttfamily arXiv:2201.07800
  [hep-ph]}}.

\bibitem{csaki2024holographicenergycorrelatorsconfining}
C.~Cs\'aki and A.~Ismail, ``{Holographic energy correlators for confining
  theories},'' \href{http://dx.doi.org/10.1007/JHEP11(2024)140}{{\em JHEP}
  {\bfseries 11} (2024) 140}, \href{http://arxiv.org/abs/2403.12123}{{\ttfamily
  arXiv:2403.12123 [hep-ph]}}.

\bibitem{CMS:2024mlf}
{\bfseries CMS} Collaboration, A.~Hayrapetyan {\em et~al.}, ``{Measurement of
  energy correlators inside jets and determination of the strong coupling
  $\alpha_\mathrm{S}(m_\mathrm{Z})$},''
  \href{http://arxiv.org/abs/2402.13864}{{\ttfamily arXiv:2402.13864
  [hep-ex]}}.

\bibitem{CMS-PAS-SMP-22-015}
{\bfseries CMS} Collaboration, ``{Measurement of energy correlators inside jets
  and determination of the strong coupling constant},'' tech. rep., CERN,
  Geneva, 2023.
\newblock \url{https://cds.cern.ch/record/2866560}.

\bibitem{alicecollaboration2024exposingpartonhadrontransitionjets}
{\bfseries ALICE} Collaboration, S.~Acharya {\em et~al.}, ``{Exposing the
  parton-hadron transition within jets with energy-energy correlators in pp
  collisions at $\sqrt{\textit s}=5.02$ TeV},''
  \href{http://arxiv.org/abs/2409.12687}{{\ttfamily arXiv:2409.12687
  [hep-ex]}}.

\bibitem{Murayama:2021xfj}
H.~Murayama, ``{Some Exact Results in QCD-like Theories},''
  \href{http://dx.doi.org/10.1103/PhysRevLett.126.251601}{{\em Phys. Rev.
  Lett.} {\bfseries 126} no.~25, (2021) 251601},
  \href{http://arxiv.org/abs/2104.01179}{{\ttfamily arXiv:2104.01179
  [hep-th]}}.

\bibitem{Csaki:2021jax}
C.~Cs\'aki, A.~Gomes, H.~Murayama, and O.~Telem, ``{Demonstration of
  Confinement and Chiral Symmetry Breaking in SO(Nc) Gauge Theories},''
  \href{http://dx.doi.org/10.1103/PhysRevLett.127.251602}{{\em Phys. Rev.
  Lett.} {\bfseries 127} no.~25, (2021) 251602},
  \href{http://arxiv.org/abs/2106.10288}{{\ttfamily arXiv:2106.10288
  [hep-th]}}.

\bibitem{Csaki:2023yas}
C.~Cs\'aki, R.~Tito~D'Agnolo, R.~S. Gupta, E.~Kuflik, T.~S. Roy, and
  M.~Ruhdorfer, ``{On the dynamical origin of the {$\eta'$} potential and the
  axion mass},'' \href{http://dx.doi.org/10.1007/JHEP10(2023)139}{{\em JHEP}
  {\bfseries 10} (2023) 139}, \href{http://arxiv.org/abs/2307.04809}{{\ttfamily
  arXiv:2307.04809 [hep-ph]}}.

\bibitem{Csaki:2024lvk}
C.~Cs\'aki, M.~Ruhdorfer, and T.~Youn, ``{Spontaneous CP Breaking in a QCD-like
  Theory},'' \href{http://arxiv.org/abs/2407.06252}{{\ttfamily arXiv:2407.06252
  [hep-ph]}}.

\bibitem{Maldacena:1997re}
J.~M. Maldacena, ``{The Large N limit of superconformal field theories and
  supergravity},'' \href{http://dx.doi.org/10.4310/ATMP.1998.v2.n2.a1}{{\em
  Adv. Theor. Math. Phys.} {\bfseries 2} (1998) 231--252},
  \href{http://arxiv.org/abs/hep-th/9711200}{{\ttfamily arXiv:hep-th/9711200}}.

\bibitem{Witten:1998qj}
E.~Witten, ``{Anti-de Sitter space and holography},''
  \href{http://dx.doi.org/10.4310/ATMP.1998.v2.n2.a2}{{\em Adv. Theor. Math.
  Phys.} {\bfseries 2} (1998) 253--291},
  \href{http://arxiv.org/abs/hep-th/9802150}{{\ttfamily arXiv:hep-th/9802150}}.

\bibitem{Gubser:1998bc}
S.~S. Gubser, I.~R. Klebanov, and A.~M. Polyakov, ``{Gauge theory correlators
  from noncritical string theory},''
  \href{http://dx.doi.org/10.1016/S0370-2693(98)00377-3}{{\em Phys. Lett. B}
  {\bfseries 428} (1998) 105--114},
  \href{http://arxiv.org/abs/hep-th/9802109}{{\ttfamily arXiv:hep-th/9802109}}.

\bibitem{Sakai:2004cn}
T.~Sakai and S.~Sugimoto, ``{Low energy hadron physics in holographic QCD},''
  \href{http://dx.doi.org/10.1143/PTP.113.843}{{\em Prog. Theor. Phys.}
  {\bfseries 113} (2005) 843--882},
  \href{http://arxiv.org/abs/hep-th/0412141}{{\ttfamily arXiv:hep-th/0412141}}.

\bibitem{deTeramond:2005su}
G.~F. de~Teramond and S.~J. Brodsky, ``{Hadronic spectrum of a holographic dual
  of QCD},'' \href{http://dx.doi.org/10.1103/PhysRevLett.94.201601}{{\em Phys.
  Rev. Lett.} {\bfseries 94} (2005) 201601},
  \href{http://arxiv.org/abs/hep-th/0501022}{{\ttfamily arXiv:hep-th/0501022}}.

\bibitem{Erlich:2005qh}
J.~Erlich, E.~Katz, D.~T. Son, and M.~A. Stephanov, ``{QCD and a holographic
  model of hadrons},''
  \href{http://dx.doi.org/10.1103/PhysRevLett.95.261602}{{\em Phys. Rev. Lett.}
  {\bfseries 95} (2005) 261602},
  \href{http://arxiv.org/abs/hep-ph/0501128}{{\ttfamily arXiv:hep-ph/0501128}}.

\bibitem{DaRold:2005mxj}
L.~Da~Rold and A.~Pomarol, ``{Chiral symmetry breaking from five dimensional
  spaces},'' \href{http://dx.doi.org/10.1016/j.nuclphysb.2005.05.009}{{\em
  Nucl. Phys. B} {\bfseries 721} (2005) 79--97},
  \href{http://arxiv.org/abs/hep-ph/0501218}{{\ttfamily arXiv:hep-ph/0501218}}.

\bibitem{Sakai:2005yt}
T.~Sakai and S.~Sugimoto, ``{More on a holographic dual of QCD},''
  \href{http://dx.doi.org/10.1143/PTP.114.1083}{{\em Prog. Theor. Phys.}
  {\bfseries 114} (2005) 1083--1118},
  \href{http://arxiv.org/abs/hep-th/0507073}{{\ttfamily arXiv:hep-th/0507073}}.

\bibitem{Karch:2006pv}
A.~Karch, E.~Katz, D.~T. Son, and M.~A. Stephanov, ``{Linear confinement and
  AdS/QCD},'' \href{http://dx.doi.org/10.1103/PhysRevD.74.015005}{{\em Phys.
  Rev. D} {\bfseries 74} (2006) 015005},
  \href{http://arxiv.org/abs/hep-ph/0602229}{{\ttfamily arXiv:hep-ph/0602229}}.

\bibitem{Csaki:2006ji}
C.~Csaki and M.~Reece, ``{Toward a systematic holographic QCD: A Braneless
  approach},'' \href{http://dx.doi.org/10.1088/1126-6708/2007/05/062}{{\em
  JHEP} {\bfseries 05} (2007) 062},
  \href{http://arxiv.org/abs/hep-ph/0608266}{{\ttfamily arXiv:hep-ph/0608266}}.

\bibitem{Csaki:2008dt}
C.~Csaki, M.~Reece, and J.~Terning, ``{The AdS/QCD Correspondence: Still
  Undelivered},'' \href{http://dx.doi.org/10.1088/1126-6708/2009/05/067}{{\em
  JHEP} {\bfseries 05} (2009) 067},
  \href{http://arxiv.org/abs/0811.3001}{{\ttfamily arXiv:0811.3001 [hep-ph]}}.

\bibitem{Erlich:2009me}
J.~Erlich, ``{How Well Does AdS/QCD Describe QCD?},''
  \href{http://dx.doi.org/10.1142/S0217751X10048718}{{\em Int. J. Mod. Phys. A}
  {\bfseries 25} (2010) 411--421},
  \href{http://arxiv.org/abs/0908.0312}{{\ttfamily arXiv:0908.0312 [hep-ph]}}.

\bibitem{Basham:1977iq}
C.~L. Basham, L.~S. Brown, S.~D. Ellis, and S.~T. Love, ``{Electron - Positron
  Annihilation Energy Pattern in Quantum Chromodynamics: Asymptotically Free
  Perturbation Theory},''
  \href{http://dx.doi.org/10.1103/PhysRevD.17.2298}{{\em Phys. Rev. D}
  {\bfseries 17} (1978) 2298}.

\bibitem{Basham:1978bw}
C.~L. Basham, L.~S. Brown, S.~D. Ellis, and S.~T. Love, ``{Energy Correlations
  in electron - Positron Annihilation: Testing QCD},''
  \href{http://dx.doi.org/10.1103/PhysRevLett.41.1585}{{\em Phys. Rev. Lett.}
  {\bfseries 41} (1978) 1585}.

\bibitem{Basham:1978zq}
C.~L. Basham, L.~S. Brown, S.~D. Ellis, and S.~T. Love, ``{Energy Correlations
  in electron-Positron Annihilation in Quantum Chromodynamics: Asymptotically
  Free Perturbation Theory},''
  \href{http://dx.doi.org/10.1103/PhysRevD.19.2018}{{\em Phys. Rev. D}
  {\bfseries 19} (1979) 2018}.

\bibitem{Basham:1979gh}
C.~L. Basham, L.~S. Brown, S.~D. Ellis, and S.~T. Love, ``{Energy Correlations
  in Perturbative Quantum Chromodynamics: A Conjecture for All Orders},''
  \href{http://dx.doi.org/10.1016/0370-2693(79)90601-4}{{\em Phys. Lett. B}
  {\bfseries 85} (1979) 297--299}.

\bibitem{Hofman:2009ug}
D.~M. Hofman, ``{Higher Derivative Gravity, Causality and Positivity of Energy
  in a UV complete QFT},''
  \href{http://dx.doi.org/10.1016/j.nuclphysb.2009.08.001}{{\em Nucl. Phys. B}
  {\bfseries 823} (2009) 174--194},
  \href{http://arxiv.org/abs/0907.1625}{{\ttfamily arXiv:0907.1625 [hep-th]}}.

\bibitem{Zhiboedov:2013opa}
A.~Zhiboedov, ``{On Conformal Field Theories With Extremal a/c Values},''
  \href{http://dx.doi.org/10.1007/JHEP04(2014)038}{{\em JHEP} {\bfseries 04}
  (2014) 038}, \href{http://arxiv.org/abs/1304.6075}{{\ttfamily arXiv:1304.6075
  [hep-th]}}.

\bibitem{Belitsky:2013xxa}
A.~V. Belitsky, S.~Hohenegger, G.~P. Korchemsky, E.~Sokatchev, and
  A.~Zhiboedov, ``{From correlation functions to event shapes},''
  \href{http://dx.doi.org/10.1016/j.nuclphysb.2014.04.020}{{\em Nucl. Phys. B}
  {\bfseries 884} (2014) 305--343},
  \href{http://arxiv.org/abs/1309.0769}{{\ttfamily arXiv:1309.0769 [hep-th]}}.

\bibitem{Belitsky:2013bja}
A.~V. Belitsky, S.~Hohenegger, G.~P. Korchemsky, E.~Sokatchev, and
  A.~Zhiboedov, ``{Event shapes in $\mathcal{N} = 4$ super-Yang-Mills
  theory},'' \href{http://dx.doi.org/10.1016/j.nuclphysb.2014.04.019}{{\em
  Nucl. Phys. B} {\bfseries 884} (2014) 206--256},
  \href{http://arxiv.org/abs/1309.1424}{{\ttfamily arXiv:1309.1424 [hep-th]}}.

\bibitem{Bousso:2015wca}
R.~Bousso, Z.~Fisher, J.~Koeller, S.~Leichenauer, and A.~C. Wall, ``{Proof of
  the Quantum Null Energy Condition},''
  \href{http://dx.doi.org/10.1103/PhysRevD.93.024017}{{\em Phys. Rev. D}
  {\bfseries 93} no.~2, (2016) 024017},
  \href{http://arxiv.org/abs/1509.02542}{{\ttfamily arXiv:1509.02542
  [hep-th]}}.

\bibitem{Faulkner:2016mzt}
T.~Faulkner, R.~G. Leigh, O.~Parrikar, and H.~Wang, ``{Modular Hamiltonians for
  Deformed Half-Spaces and the Averaged Null Energy Condition},''
  \href{http://dx.doi.org/10.1007/JHEP09(2016)038}{{\em JHEP} {\bfseries 09}
  (2016) 038}, \href{http://arxiv.org/abs/1605.08072}{{\ttfamily
  arXiv:1605.08072 [hep-th]}}.

\bibitem{Bousso:2016vlt}
R.~Bousso, ``{Asymptotic Entropy Bounds},''
  \href{http://dx.doi.org/10.1103/PhysRevD.94.024018}{{\em Phys. Rev. D}
  {\bfseries 94} no.~2, (2016) 024018},
  \href{http://arxiv.org/abs/1606.02297}{{\ttfamily arXiv:1606.02297
  [hep-th]}}.

\bibitem{Hartman:2016lgu}
T.~Hartman, S.~Kundu, and A.~Tajdini, ``{Averaged Null Energy Condition from
  Causality},'' \href{http://dx.doi.org/10.1007/JHEP07(2017)066}{{\em JHEP}
  {\bfseries 07} (2017) 066}, \href{http://arxiv.org/abs/1610.05308}{{\ttfamily
  arXiv:1610.05308 [hep-th]}}.

\bibitem{Casini:2017roe}
H.~Casini, E.~Teste, and G.~Torroba, ``{Modular Hamiltonians on the null plane
  and the Markov property of the vacuum state},''
  \href{http://dx.doi.org/10.1088/1751-8121/aa7eaa}{{\em J. Phys. A} {\bfseries
  50} no.~36, (2017) 364001}, \href{http://arxiv.org/abs/1703.10656}{{\ttfamily
  arXiv:1703.10656 [hep-th]}}.

\bibitem{Balakrishnan:2017bjg}
S.~Balakrishnan, T.~Faulkner, Z.~U. Khandker, and H.~Wang, ``{A General Proof
  of the Quantum Null Energy Condition},''
  \href{http://dx.doi.org/10.1007/JHEP09(2019)020}{{\em JHEP} {\bfseries 09}
  (2019) 020}, \href{http://arxiv.org/abs/1706.09432}{{\ttfamily
  arXiv:1706.09432 [hep-th]}}.

\bibitem{Cordova:2017zej}
C.~Cordova, J.~Maldacena, and G.~J. Turiaci, ``{Bounds on OPE Coefficients from
  Interference Effects in the Conformal Collider},''
  \href{http://dx.doi.org/10.1007/JHEP11(2017)032}{{\em JHEP} {\bfseries 11}
  (2017) 032}, \href{http://arxiv.org/abs/1710.03199}{{\ttfamily
  arXiv:1710.03199 [hep-th]}}.

\bibitem{Cordova:2017dhq}
C.~Cordova and K.~Diab, ``{Universal Bounds on Operator Dimensions from the
  Average Null Energy Condition},''
  \href{http://dx.doi.org/10.1007/JHEP02(2018)131}{{\em JHEP} {\bfseries 02}
  (2018) 131}, \href{http://arxiv.org/abs/1712.01089}{{\ttfamily
  arXiv:1712.01089 [hep-th]}}.

\bibitem{Leichenauer:2018obf}
S.~Leichenauer, A.~Levine, and A.~Shahbazi-Moghaddam, ``{Energy density from
  second shape variations of the von Neumann entropy},''
  \href{http://dx.doi.org/10.1103/PhysRevD.98.086013}{{\em Phys. Rev. D}
  {\bfseries 98} no.~8, (2018) 086013},
  \href{http://arxiv.org/abs/1802.02584}{{\ttfamily arXiv:1802.02584
  [hep-th]}}.

\bibitem{Kravchuk:2018htv}
P.~Kravchuk and D.~Simmons-Duffin, ``{Light-ray operators in conformal field
  theory},'' \href{http://dx.doi.org/10.1007/JHEP11(2018)102}{{\em JHEP}
  {\bfseries 11} (2018) 102}, \href{http://arxiv.org/abs/1805.00098}{{\ttfamily
  arXiv:1805.00098 [hep-th]}}.

\bibitem{Cordova:2018ygx}
C.~C\'ordova and S.-H. Shao, ``{Light-ray Operators and the BMS Algebra},''
  \href{http://dx.doi.org/10.1103/PhysRevD.98.125015}{{\em Phys. Rev. D}
  {\bfseries 98} no.~12, (2018) 125015},
  \href{http://arxiv.org/abs/1810.05706}{{\ttfamily arXiv:1810.05706
  [hep-th]}}.

\bibitem{Ceyhan:2018zfg}
F.~Ceyhan and T.~Faulkner, ``{Recovering the QNEC from the ANEC},''
  \href{http://dx.doi.org/10.1007/s00220-020-03751-y}{{\em Commun. Math. Phys.}
  {\bfseries 377} no.~2, (2020) 999--1045},
  \href{http://arxiv.org/abs/1812.04683}{{\ttfamily arXiv:1812.04683
  [hep-th]}}.

\bibitem{Manenti:2019kbl}
A.~Manenti, A.~Stergiou, and A.~Vichi, ``{Implications of ANEC for SCFTs in
  four dimensions},'' \href{http://dx.doi.org/10.1007/JHEP01(2020)093}{{\em
  JHEP} {\bfseries 01} (2020) 093},
  \href{http://arxiv.org/abs/1905.09293}{{\ttfamily arXiv:1905.09293
  [hep-th]}}.

\bibitem{Balakrishnan:2019gxl}
S.~Balakrishnan, V.~Chandrasekaran, T.~Faulkner, A.~Levine, and
  A.~Shahbazi-Moghaddam, ``{Entropy variations and light ray operators from
  replica defects},'' \href{http://dx.doi.org/10.1007/JHEP09(2022)217}{{\em
  JHEP} {\bfseries 09} (2022) 217},
  \href{http://arxiv.org/abs/1906.08274}{{\ttfamily arXiv:1906.08274
  [hep-th]}}.

\bibitem{Belin:2020lsr}
A.~Belin, D.~M. Hofman, G.~Mathys, and M.~T. Walters, ``{On the stress tensor
  light-ray operator algebra},''
  \href{http://dx.doi.org/10.1007/JHEP05(2021)033}{{\em JHEP} {\bfseries 05}
  (2021) 033}, \href{http://arxiv.org/abs/2011.13862}{{\ttfamily
  arXiv:2011.13862 [hep-th]}}.

\bibitem{Korchemsky:2021htm}
G.~P. Korchemsky and A.~Zhiboedov, ``{On the light-ray algebra in conformal
  field theories},'' \href{http://dx.doi.org/10.1007/JHEP02(2022)140}{{\em
  JHEP} {\bfseries 02} (2022) 140},
  \href{http://arxiv.org/abs/2109.13269}{{\ttfamily arXiv:2109.13269
  [hep-th]}}.

\bibitem{Hartman:2023qdn}
T.~Hartman and G.~Mathys, ``{Averaged null energy and the renormalization
  group},'' \href{http://dx.doi.org/10.1007/JHEP12(2023)139}{{\em JHEP}
  {\bfseries 12} (2023) 139}, \href{http://arxiv.org/abs/2309.14409}{{\ttfamily
  arXiv:2309.14409 [hep-th]}}.

\bibitem{Hartman:2024xkw}
T.~Hartman and G.~Mathys, ``{Light-ray sum rules and the c-anomaly},''
  \href{http://dx.doi.org/10.1007/JHEP08(2024)008}{{\em JHEP} {\bfseries 08}
  (2024) 008}, \href{http://arxiv.org/abs/2405.10137}{{\ttfamily
  arXiv:2405.10137 [hep-th]}}.

\bibitem{Dixon:2019uzg}
L.~J. Dixon, I.~Moult, and H.~X. Zhu, ``{Collinear limit of the energy-energy
  correlator},'' \href{http://dx.doi.org/10.1103/PhysRevD.100.014009}{{\em
  Phys. Rev. D} {\bfseries 100} no.~1, (2019) 014009},
  \href{http://arxiv.org/abs/1905.01310}{{\ttfamily arXiv:1905.01310
  [hep-ph]}}.

\bibitem{Chen:2019bpb}
H.~Chen, M.-X. Luo, I.~Moult, T.-Z. Yang, X.~Zhang, and H.~X. Zhu, ``{Three
  point energy correlators in the collinear limit: symmetries, dualities and
  analytic results},'' \href{http://dx.doi.org/10.1007/JHEP08(2020)028}{{\em
  JHEP} {\bfseries 08} no.~08, (2020) 028},
  \href{http://arxiv.org/abs/1912.11050}{{\ttfamily arXiv:1912.11050
  [hep-ph]}}.

\bibitem{Chen:2020vvp}
H.~Chen, I.~Moult, X.~Zhang, and H.~X. Zhu, ``{Rethinking jets with energy
  correlators: Tracks, resummation, and analytic continuation},''
  \href{http://dx.doi.org/10.1103/PhysRevD.102.054012}{{\em Phys. Rev. D}
  {\bfseries 102} no.~5, (2020) 054012},
  \href{http://arxiv.org/abs/2004.11381}{{\ttfamily arXiv:2004.11381
  [hep-ph]}}.

\bibitem{Chen:2020adz}
H.~Chen, I.~Moult, and H.~X. Zhu, ``{Quantum Interference in Jet Substructure
  from Spinning Gluons},''
  \href{http://dx.doi.org/10.1103/PhysRevLett.126.112003}{{\em Phys. Rev.
  Lett.} {\bfseries 126} no.~11, (2021) 112003},
  \href{http://arxiv.org/abs/2011.02492}{{\ttfamily arXiv:2011.02492
  [hep-ph]}}.

\bibitem{Chen:2021gdk}
H.~Chen, I.~Moult, and H.~X. Zhu, ``{Spinning gluons from the QCD light-ray
  OPE},'' \href{http://dx.doi.org/10.1007/JHEP08(2022)233}{{\em JHEP}
  {\bfseries 08} (2022) 233}, \href{http://arxiv.org/abs/2104.00009}{{\ttfamily
  arXiv:2104.00009 [hep-ph]}}.

\bibitem{Holguin:2022epo}
J.~Holguin, I.~Moult, A.~Pathak, and M.~Procura, ``{New paradigm for precision
  top physics: Weighing the top with energy correlators},''
  \href{http://dx.doi.org/10.1103/PhysRevD.107.114002}{{\em Phys. Rev. D}
  {\bfseries 107} no.~11, (2023) 114002},
  \href{http://arxiv.org/abs/2201.08393}{{\ttfamily arXiv:2201.08393
  [hep-ph]}}.

\bibitem{Chen:2022jhb}
H.~Chen, I.~Moult, J.~Sandor, and H.~X. Zhu, ``{Celestial blocks and transverse
  spin in the three-point energy correlator},''
  \href{http://dx.doi.org/10.1007/JHEP09(2022)199}{{\em JHEP} {\bfseries 09}
  (2022) 199}, \href{http://arxiv.org/abs/2202.04085}{{\ttfamily
  arXiv:2202.04085 [hep-ph]}}.

\bibitem{Chen:2022swd}
H.~Chen, I.~Moult, J.~Thaler, and H.~X. Zhu, ``{Non-Gaussianities in collider
  energy flux},'' \href{http://dx.doi.org/10.1007/JHEP07(2022)146}{{\em JHEP}
  {\bfseries 07} (2022) 146}, \href{http://arxiv.org/abs/2205.02857}{{\ttfamily
  arXiv:2205.02857 [hep-ph]}}.

\bibitem{Lee:2022ige}
K.~Lee, B.~Me\c{c}aj, and I.~Moult, ``{Conformal Colliders Meet the LHC},''
  \href{http://arxiv.org/abs/2205.03414}{{\ttfamily arXiv:2205.03414
  [hep-ph]}}.

\bibitem{Bossi:2024qho}
H.~Bossi, A.~S. Kudinoor, I.~Moult, D.~Pablos, A.~Rai, and K.~Rajagopal,
  ``{Imaging the Wakes of Jets with Energy-Energy-Energy Correlators},''
  \href{http://arxiv.org/abs/2407.13818}{{\ttfamily arXiv:2407.13818
  [hep-ph]}}.

\bibitem{Lee:2024jnt}
K.~Lee, F.~Turro, and X.~Yao, ``{Quantum Computing for Energy Correlators},''
  \href{http://arxiv.org/abs/2409.13830}{{\ttfamily arXiv:2409.13830
  [hep-ph]}}.

\bibitem{Lee:2024esz}
K.~Lee, A.~Pathak, I.~Stewart, and Z.~Sun, ``{Nonperturbative Effects in Energy
  Correlators: From Characterizing Confinement Transition to Improving
  $\alpha_s$ Extraction},'' \href{http://arxiv.org/abs/2405.19396}{{\ttfamily
  arXiv:2405.19396 [hep-ph]}}.

\bibitem{Chen:2024nyc}
H.~Chen, P.~F. Monni, Z.~Xu, and H.~X. Zhu, ``{Scaling violation in power
  corrections to energy correlators from the light-ray OPE},''
  \href{http://arxiv.org/abs/2406.06668}{{\ttfamily arXiv:2406.06668
  [hep-ph]}}.

\bibitem{Liu:2024lxy}
X.~Liu, W.~Vogelsang, F.~Yuan, and H.~X. Zhu, ``{Universality in the Near-Side
  Energy-Energy Correlator},''
  \href{http://arxiv.org/abs/2410.16371}{{\ttfamily arXiv:2410.16371
  [hep-ph]}}.

\bibitem{Chen:2024nfl}
A.-P. Chen, X.~Liu, and Y.-Q. Ma, ``{Shedding Light on Hadronization by
  Quarkonium Energy Correlator},''
  \href{http://dx.doi.org/10.1103/PhysRevLett.133.191901}{{\em Phys. Rev.
  Lett.} {\bfseries 133} (2024) 19},
  \href{http://arxiv.org/abs/2405.10056}{{\ttfamily arXiv:2405.10056
  [hep-ph]}}.

\bibitem{ameenthesis}
A.~Ismail, {\em {Mr. Dilaton's Adventures in Phenomenology}}.
\newblock PhD thesis, Cornell U., 2024.

\bibitem{Randall_1999}
L.~Randall and R.~Sundrum, ``{A Large mass hierarchy from a small extra
  dimension},'' \href{http://dx.doi.org/10.1103/PhysRevLett.83.3370}{{\em Phys.
  Rev. Lett.} {\bfseries 83} (1999) 3370--3373},
  \href{http://arxiv.org/abs/hep-ph/9905221}{{\ttfamily arXiv:hep-ph/9905221}}.

\bibitem{Dixon_2018}
L.~J. Dixon, M.-X. Luo, V.~Shtabovenko, T.-Z. Yang, and H.~X. Zhu,
  ``{Analytical Computation of Energy-Energy Correlation at Next-to-Leading
  Order in QCD},'' \href{http://dx.doi.org/10.1103/PhysRevLett.120.102001}{{\em
  Phys. Rev. Lett.} {\bfseries 120} no.~10, (2018) 102001},
  \href{http://arxiv.org/abs/1801.03219}{{\ttfamily arXiv:1801.03219
  [hep-ph]}}.

\bibitem{Csaki:2021gfm}
C.~Cs\'aki, S.~Hong, G.~Kurup, S.~J. Lee, M.~Perelstein, and W.~Xue,
  ``{Continuum dark matter},''
  \href{http://dx.doi.org/10.1103/PhysRevD.105.035025}{{\em Phys. Rev. D}
  {\bfseries 105} no.~3, (2022) 035025},
  \href{http://arxiv.org/abs/2105.07035}{{\ttfamily arXiv:2105.07035
  [hep-ph]}}.

\bibitem{Csaki:2021xpy}
C.~Cs\'aki, S.~Hong, G.~Kurup, S.~J. Lee, M.~Perelstein, and W.~Xue,
  ``{Z-Portal Continuum Dark Matter},''
  \href{http://dx.doi.org/10.1103/PhysRevLett.128.081807}{{\em Phys. Rev.
  Lett.} {\bfseries 128} no.~8, (2022) 081807},
  \href{http://arxiv.org/abs/2105.14023}{{\ttfamily arXiv:2105.14023
  [hep-ph]}}.

\bibitem{Csaki:2022lnq}
C.~Csaki, A.~Ismail, and S.~J. Lee, ``{The continuum dark matter zoo},''
  \href{http://dx.doi.org/10.1007/JHEP02(2023)053}{{\em JHEP} {\bfseries 02}
  (2023) 053}, \href{http://arxiv.org/abs/2210.16326}{{\ttfamily
  arXiv:2210.16326 [hep-ph]}}.

\bibitem{Ferrante:2023fpx}
S.~Ferrante, S.~J. Lee, and M.~Perelstein, ``{Collider signatures of
  near-continuum dark matter},''
  \href{http://dx.doi.org/10.1007/JHEP05(2024)215}{{\em JHEP} {\bfseries 05}
  (2024) 215}, \href{http://arxiv.org/abs/2306.13009}{{\ttfamily
  arXiv:2306.13009 [hep-ph]}}.

\bibitem{Katz:2015zba}
A.~Katz, M.~Reece, and A.~Sajjad, ``{Continuum-mediated dark
  matter\textendash{}baryon scattering},''
  \href{http://dx.doi.org/10.1016/j.dark.2016.01.002}{{\em Phys. Dark Univ.}
  {\bfseries 12} (2016) 24--36},
  \href{http://arxiv.org/abs/1509.03628}{{\ttfamily arXiv:1509.03628
  [hep-ph]}}.

\bibitem{Chaffey:2021tmj}
I.~Chaffey, S.~Fichet, and P.~Tanedo, ``{Continuum-Mediated Self-Interacting
  Dark Matter},'' \href{http://dx.doi.org/10.1007/JHEP06(2021)008}{{\em JHEP}
  {\bfseries 06} (2021) 008}, \href{http://arxiv.org/abs/2102.05674}{{\ttfamily
  arXiv:2102.05674 [hep-ph]}}.

\bibitem{Falkowski:2008fz}
A.~Falkowski and M.~Perez-Victoria, ``{Electroweak Breaking on a Soft Wall},''
  \href{http://dx.doi.org/10.1088/1126-6708/2008/12/107}{{\em JHEP} {\bfseries
  12} (2008) 107}, \href{http://arxiv.org/abs/0806.1737}{{\ttfamily
  arXiv:0806.1737 [hep-ph]}}.

\bibitem{Bellazzini:2015cgj}
B.~Bellazzini, C.~Cs\'aki, J.~Hubisz, S.~J. Lee, J.~Serra, and J.~Terning,
  ``{Quantum Critical Higgs},''
  \href{http://dx.doi.org/10.1103/PhysRevX.6.041050}{{\em Phys. Rev. X}
  {\bfseries 6} no.~4, (2016) 041050},
  \href{http://arxiv.org/abs/1511.08218}{{\ttfamily arXiv:1511.08218
  [hep-ph]}}.

\bibitem{Csaki:2018kxb}
C.~Cs\'aki, G.~Lee, S.~J. Lee, S.~Lombardo, and O.~Telem, ``{Continuum
  Naturalness},'' \href{http://dx.doi.org/10.1007/JHEP03(2019)142}{{\em JHEP}
  {\bfseries 03} (2019) 142}, \href{http://arxiv.org/abs/1811.06019}{{\ttfamily
  arXiv:1811.06019 [hep-ph]}}.

\bibitem{Cabrer:2009we}
J.~A. Cabrer, G.~von Gersdorff, and M.~Quiros, ``{Soft-Wall Stabilization},''
  \href{http://dx.doi.org/10.1088/1367-2630/12/7/075012}{{\em New J. Phys.}
  {\bfseries 12} (2010) 075012},
  \href{http://arxiv.org/abs/0907.5361}{{\ttfamily arXiv:0907.5361 [hep-ph]}}.

\bibitem{Megias:2019vdb}
E.~Meg\'\i{}as and M.~Quir\'os, ``{Gapped Continuum Kaluza-Klein spectrum},''
  \href{http://dx.doi.org/10.1007/JHEP08(2019)166}{{\em JHEP} {\bfseries 08}
  (2019) 166}, \href{http://arxiv.org/abs/1905.07364}{{\ttfamily
  arXiv:1905.07364 [hep-ph]}}.

\bibitem{Megias:2021mgj}
E.~Megias and M.~Quiros, ``{The Continuum Linear Dilaton},''
  \href{http://dx.doi.org/10.5506/APhysPolB.52.711}{{\em Acta Phys. Polon. B}
  {\bfseries 52} no.~6-7, (2021) 711},
  \href{http://arxiv.org/abs/2104.10260}{{\ttfamily arXiv:2104.10260
  [hep-ph]}}.

\bibitem{Cai:2009ax}
H.~Cai, H.-C. Cheng, A.~D. Medina, and J.~Terning, ``{Continuum Superpartners
  from Supersymmetric Unparticles},''
  \href{http://dx.doi.org/10.1103/PhysRevD.80.115009}{{\em Phys. Rev. D}
  {\bfseries 80} (2009) 115009},
  \href{http://arxiv.org/abs/0910.3925}{{\ttfamily arXiv:0910.3925 [hep-ph]}}.

\bibitem{Cai:2011ww}
H.~Cai, H.-C. Cheng, A.~D. Medina, and J.~Terning, ``{SUSY Hidden in the
  Continuum},'' \href{http://dx.doi.org/10.1103/PhysRevD.85.015019}{{\em Phys.
  Rev. D} {\bfseries 85} (2012) 015019},
  \href{http://arxiv.org/abs/1108.3574}{{\ttfamily arXiv:1108.3574 [hep-ph]}}.

\bibitem{Gao:2019gfw}
C.~Gao, A.~Shayegan~Shirazi, and J.~Terning, ``{Collider Phenomenology of a
  Gluino Continuum},'' \href{http://dx.doi.org/10.1007/JHEP01(2020)102}{{\em
  JHEP} {\bfseries 01} (2020) 102},
  \href{http://arxiv.org/abs/1909.04061}{{\ttfamily arXiv:1909.04061
  [hep-ph]}}.

\bibitem{Aoki:2023tjm}
S.~Aoki, ``{Continuous spectrum on cosmological collider},''
  \href{http://dx.doi.org/10.1088/1475-7516/2023/04/002}{{\em JCAP} {\bfseries
  04} (2023) 002}, \href{http://arxiv.org/abs/2301.07920}{{\ttfamily
  arXiv:2301.07920 [hep-th]}}.

\bibitem{Hubisz:2024xnj}
J.~Hubisz, S.~J. Lee, H.~Li, and B.~Sambasivam, ``{Cosmological Quasiparticles
  and the Cosmological Collider},''
  \href{http://arxiv.org/abs/2408.08951}{{\ttfamily arXiv:2408.08951
  [astro-ph.CO]}}.

\bibitem{MR493193}
R.~Wong, ``Asymptotic expansions of {H}ankel transforms of functions with
  logarithmic singularities,''
  \href{http://dx.doi.org/10.1016/0898-1221(77)90084-0}{{\em Comput. Math.
  Appl.} {\bfseries 3} no.~4, (1977) 271--286}.

\bibitem{MR804942}
C.~L. Frenzen and R.~Wong, ``A note on asymptotic evaluation of some {H}ankel
  transforms,'' \href{http://dx.doi.org/10.2307/2008143}{{\em Math. Comp.}
  {\bfseries 45} no.~172, (1985) 537--548}.

\bibitem{MR811192}
B.~Gabutti, ``An asymptotic approximation for a class of oscillatory infinite
  integrals,'' \href{http://dx.doi.org/10.1137/0722072}{{\em SIAM J. Numer.
  Anal.} {\bfseries 22} no.~6, (1985) 1191--1199}.

\bibitem{MR903988}
B.~Gabutti and P.~Lepora, ``A novel approach for the determination of
  asymptotic expansions of certain oscillatory integrals,''
  \href{http://dx.doi.org/10.1016/0377-0427(87)90189-0}{{\em J. Comput. Appl.
  Math.} {\bfseries 19} no.~2, (1987) 189--206}.

\bibitem{MR1851050}
R.~Wong, \href{http://dx.doi.org/10.1137/1.9780898719260}{{\em Asymptotic
  approximations of integrals}}, vol.~34 of {\em Classics in Applied
  Mathematics}.
\newblock Society for Industrial and Applied Mathematics (SIAM), Philadelphia,
  PA, 2001.

\bibitem{MR3149389}
E.~A. Galapon and K.~M.~L. Martinez, ``Exactification of the {P}oincar\'e{}
  asymptotic expansion of the {H}ankel integral: spectacularly accurate
  asymptotic expansions and non-asymptotic scales,''
  \href{http://dx.doi.org/10.1098/rspa.2013.0529}{{\em Proc. R. Soc. Lond. Ser.
  A Math. Phys. Eng. Sci.} {\bfseries 470} no.~2162, (2014) 20130529, 16}.

\bibitem{de_Florian_2005}
D.~de~Florian and M.~Grazzini, ``{The Back-to-back region in e+ e-
  energy-energy correlation},''
  \href{http://dx.doi.org/10.1016/j.nuclphysb.2004.10.051}{{\em Nucl. Phys. B}
  {\bfseries 704} (2005) 387--403},
  \href{http://arxiv.org/abs/hep-ph/0407241}{{\ttfamily arXiv:hep-ph/0407241}}.

\bibitem{futurework}
C.~Cs\'aki, A.~Ismail, and L.~Kiriliuk, ``{Holographic energy correlators with
  asymptotic freedom},'' {\em in preparation\hspace{-0.1cm}} .

\end{thebibliography}\endgroup
\end{document}